\documentclass[fleqn,10pt]{wlscirep}
\usepackage[utf8]{inputenc}
\usepackage[T1]{fontenc}
\usepackage{braket}
\usepackage{xcolor}
\usepackage{graphicx}
\usepackage{dcolumn}
\usepackage{bm}
\usepackage{filecontents}
\title{Phase estimation algorithm for the multibeam optical metrology}

\author[1,2]{V.\,V.\,Zemlyanov}
\author[1,2,3,4]{N.\,S.\,Kirsanov}
\author[1,2,4]{M.\,R.\,Perelshtein}
\author[2]{D.\,I.\,Lykov}
\author[5,1,2]{O.\,V.\,Misochko}
\author[5,1,2]{M.\,V.\,Lebedev}
\author[6,3,*]{V.\,M.\,Vinokur}
\author[1,2]{G.\,B.\,Lesovik}
\affil[1]{Terra Quantum AG, St.\,Gallerstrasse 16A, 9400 Rorschach, Switzerland.}
\affil[2]{Moscow Institute of Physics and Technology, 141700, Institutskii Per. 9, Dolgoprudny, Moscow Distr., Russian Federation}
\affil[3]{Consortium for Advanced Science and Engineering (CASE), University of Chicago, 5801 S Ellis Ave, Chicago, IL 60637, USA}
\affil[4]{Low Temperature Laboratory, Department of Applied Physics, Aalto University, P.O. Box 15100, FI-00076 AALTO, Finland}
\affil[5]{Institute of Solid State Physics, Russian Academy of Sciences, 142432, Chernogolovka, Moscow Distr., Russian Federation}
\affil[6]{Materials Science Division, Argonne
National Laboratory, 9700 S. Cass Ave., Argonne, IL 60439, USA}

\affil[*]{vinokour@anl.gov}

\begin{abstract}
Unitary Fourier transform lies at the core of the multitudinous computational and metrological algorithms.
Here we show experimentally how the unitary Fourier transform-based phase estimation protocol, used namely in quantum metrology, can be translated into the classical linear optical framework.
The developed setup made of beam splitters, mirrors and phase shifters demonstrates how the classical coherence, similarly to the quantum coherence, poses a resource for obtaining information about the measurable physical quantities.
Our study opens route to the reliable implementation of the  small-scale unitary algorithms on path-encoded qudits, thus establishing an easily accessible platform for unitary computation.
\end{abstract}
\begin{document}

\flushbottom
\maketitle
%
%
\thispagestyle{empty}


\section*{Introduction}


Unitary Fourier transform is a quintessential component for a multitude of quantum computational algorithms\,\cite{Shor1994, Lloyd2007, Peruzzo2014} as it underlies a versatile phase estimation routine\,\cite{NielsenChuang} which is at the core of various quantum metrological protocols\,\cite{Giovannetti2004, Lesovik2010, Suslov2011, Giovannetti2011}.
Such phase-sensitive protocols, utilizing coherence for measurements of physical quantities, find use in quantum sensors\,\cite{Degen2017, Pirandola2018}, notably in the qudit-based devices (e.g., based on the superconducting artificial atoms or NV centers) for determining magnetic and electric fields\,\cite{Waldherr2011, Bal2012, Puentes2014, Bonato2015, Chen2017, Danilin, Shlyakhov}.
Importantly, since these protocols do not necessarily employ quantum entanglement\,\cite{Higgins2007}, they may be implemented on the systems that manifest wave yet classical behavior.
Therefore, methods borrowed from quantum metrology can be applied to the classical optical phase measurements\,\cite{Knill, Poland, Dowling, Carolan, Tan2019}, which, in particular, can be used to measure the position, velocity, and displacement of physical objects.
Here we report on constructing a complex linear-optic-based device capable to carry out the Fourier-based phase estimation algorithm.
The metrological potential of the intricate multiple-beam interference schemes can be, for instance, seen in the LIGO optical gravitational wave detector\,\cite{LIGO} where the Heisenberg-limited sensitivity is achieved through combining Michelson and Fabry-P\'erot interferometers and employing the squeezed states of light.
%

Our approach is predicated upon the fact that any finite-dimensional unitary matrix can be realized by means of 50:50 beam splitters, phase shifters and mirrors\,\cite{Reck}. 
In order to better demonstrate the computational capabilities of the linear optics, we adopt the laser as a source of the light having the coherence length by far exceeding the size of the setup. 
This ensures the speed of measurements that is sufficient to support the stability of the interference pattern during the time necessary for collecting the required statistics. 
Note that in the single-photon regime, the time needed to obtain the same statistics would be much too long to preserve the same quality of the interference pattern throughout the entire measurement procedure.
Using the multiphoton source does not eliminate the unitary nature of the algorithm, which employs for this moderate computation scale only the wave aspect of the signal. 
Switching to the single-photon source for practical computation purposes will translate the scheme into the fully quantum one, while maintaining the major characteristics manifested by the present device.
A general architecture for such a multiport interferometer was first proposed by Reck \textit{et al}\,\cite{Reck} and then further reframed by Clements \textit{et al}\,\cite{Clements}.
The theoretical prospects of the proposed architecture were discussed in Refs.\,[\citeonline{Qi2018,Guise}]. 
Experimentally, it was shown that linear optical protocols can be implemented on a photonic chip\,\cite{Carolan,Harris}. 
Yet, the practical engineering of such a structure remains highly challenging.

In what follows, we will overview our algorithm and the theoretical background, describe our experimental layout, and construct the analytical description of the computational scheme.
Finally, we discuss the results and outline the future research directions.
\begin{figure}[t]
    \noindent\centering{
    \includegraphics[width=100mm]{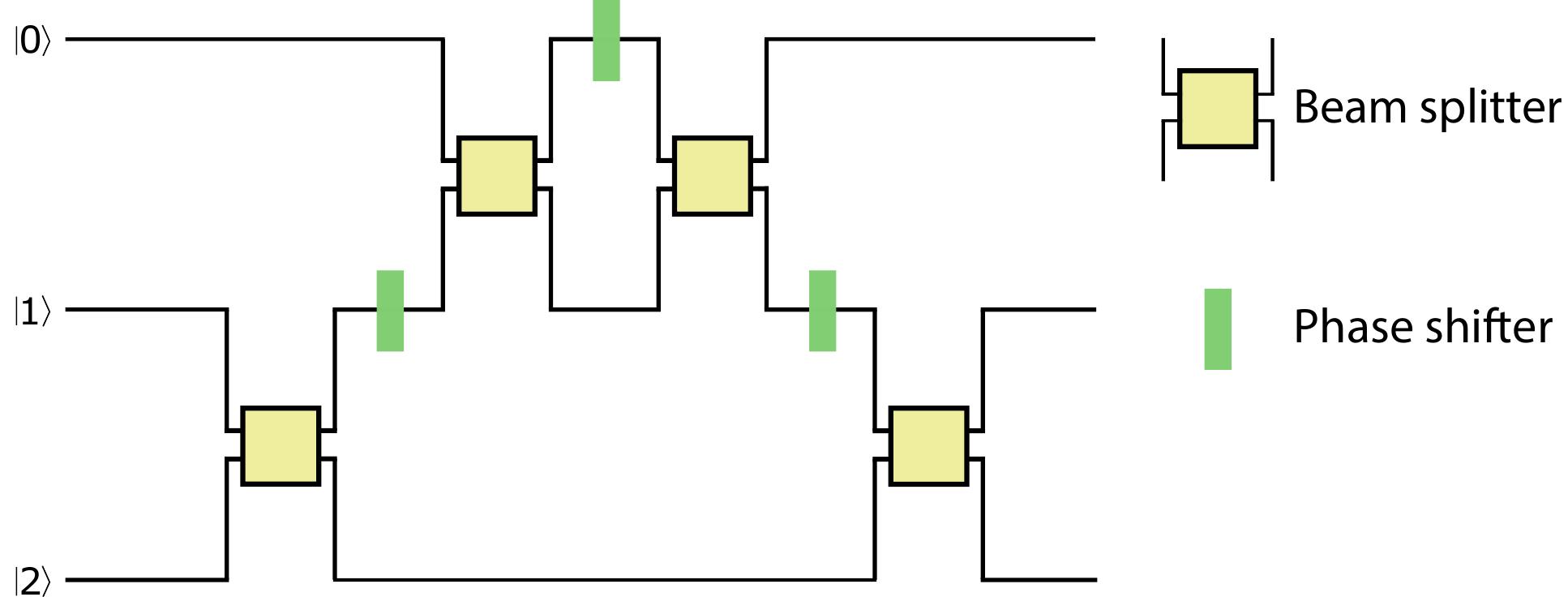}
    }
    \caption{Optical circuit realizing the qutrit quantum Fourier transformation.}
    \label{qutrit_circ}
\end{figure}

\section*{Preliminaries}
\label{theory}
\subsection*{Algorithm description}
We start with the description of the Fourier phase-estimation algorithm operating in the qudit regime. The initial qudit state is taken as a superposition of all computational states:
\begin{equation}
\label{initial_state}
    \ket{\Psi_\phi} = \frac{1}{\sqrt{d}}\sum_{k=0}^{d-1}{e^{i k \phi}\ket{k}},
\end{equation}
where $\{\ket k\}_{k=0}^{d-1}$ is an orthonormal computational basis in the qudit's Hilbert space.  Additionally, we let $\phi = \frac{2m \pi}{d}$,  $m \in \{0,1,\dots,d-1\}$.
The algorithm has to unambiguously determine the value of $\phi$ via a single-shot measurement of the qudit state. 
This is achieved by applying a base-$d$ quantum Fourier transformation with the corresponding unitary operator $\hat F$,
\begin{equation}
\label{Fourier}
    \hat{F} \ket{n} = \frac{1}{\sqrt{d}}\sum_{k=0}^{d-1}{e^{-2 \pi i n k / d}\ket{k}}.
\end{equation}
\begin{figure*}[t]
    \noindent\centering{
    \includegraphics[width=150mm]{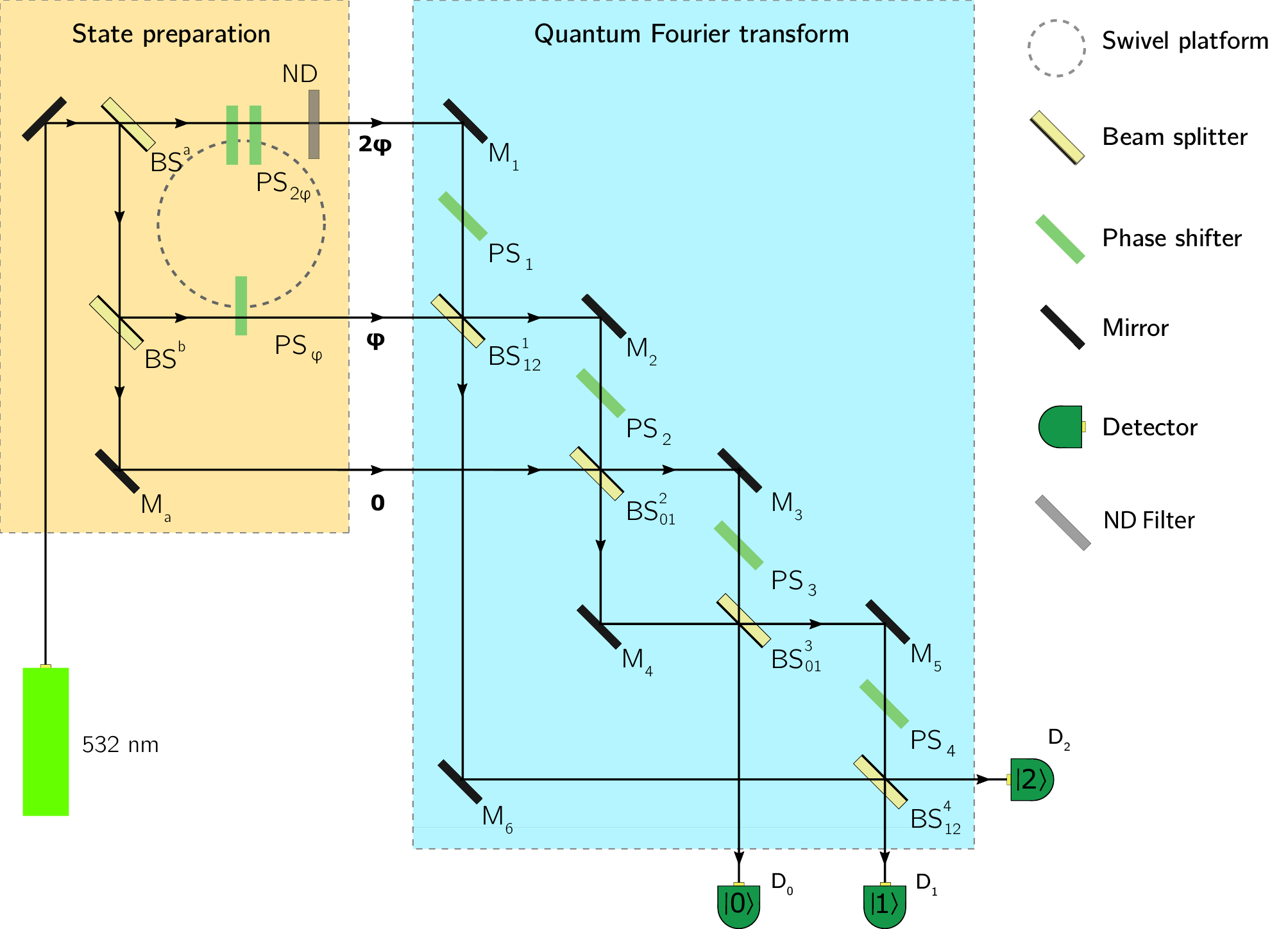}
    }
    \caption{
        Experimental scheme for the qutrit case of the metrological algorithm.
    }
    \label{scheme}
\end{figure*}
The action of $\hat F$ on the initial state $\ket{\Psi_\phi}$ yields one of the states from the computational set $\{\ket k\}_{k=0}^{d-1}$ depending on $\phi = \frac{2 \pi m}{d}$:
\begin{equation}
    \ket{\Psi_\text{out}} = \hat{F} \ket{\Psi_\phi} = \ket{m}.
\end{equation}
Accordingly, by measuring the output state $\ket{\Psi_\text{out}}$ one determines the value of $\phi$.

The above algorithm appears as a subroutine in a family of conditional sequential sensing protocols with the scaling corresponding to the Heisenberg limit, for example, the Kitaev protocol.
An essential principle of these protocols is the phase encoding: on each step of the procedure, the state of the qudit is tagged with the phase $\phi$ (as in Eq.\,(\ref{initial_state})) which depends on the unknown constant physical value to be determined and on the sensing period of the step $t$.

\subsection*{Optical scheme}
Now we introduce our optical framework. 
In this setting the qudit is represented by the $d$ coherent beams. 
Each element of its $d$-dimensional state vector is a complex amplitude of the corresponding beam. 
Accordingly, the state vector transforms when the light passes through the arrangement of beam splitters, phase shifters and mirrors.
The task of constructing a particular unitary operator reduces to its decomposition into a sequence of the two-dimensional beam splitter transformations and individual phase shifts. 
In this section we devise base-3 (qutrit) scheme to carry out the Fourier transformation
\begin{equation}
    \hat{F}=\frac{1}{\sqrt{3}}\begin{pmatrix} 
                1 & 1 & 1 \\
                1 & e^{4\pi i/3} & e^{2\pi i/3} \\
                1 & e^{2\pi i/3} & e^{4\pi i/3}
            \end{pmatrix}.
\end{equation}
\begin{figure*}[t]
    \noindent\centering{
    \includegraphics[width=180mm]{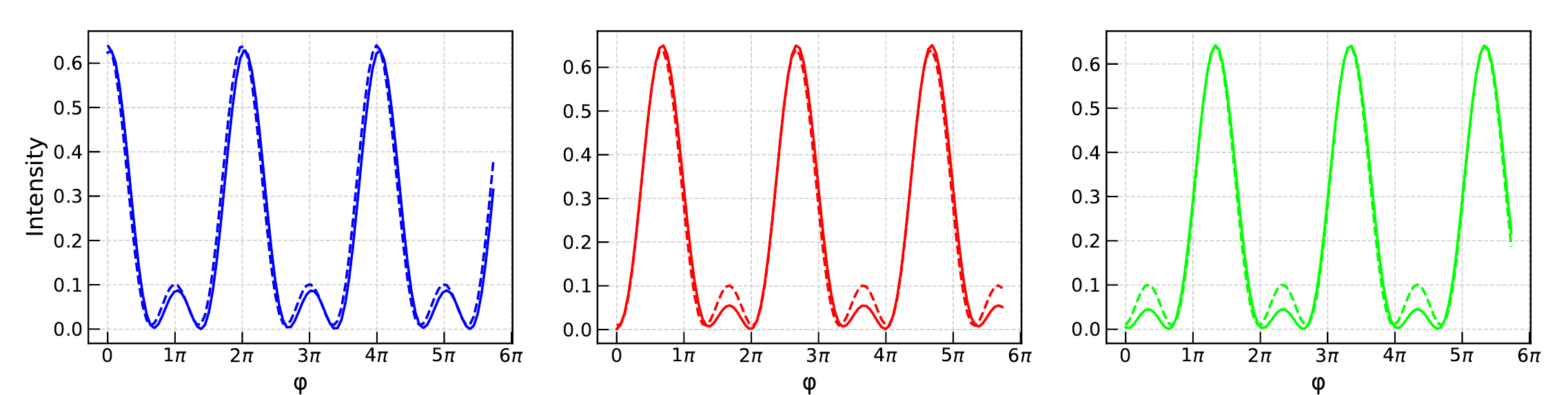}
    }
    \caption{Theoretical plots of the intensity on the $\ket 0$ (blue line), $\ket 1$ (red line) and $\ket 2$ (green line) detectors as functions of $\phi$. Dashed line shows the results obtained by means of Eq.\,(\ref{susq3}), whereas the solid line refers to Eq.\,(\ref{susq3_fixed}).}
    \label{signals_theory}
\end{figure*}
The matrix $\hat A^\chi_{jk}(\alpha,\theta)$ of an arbitrary lossless beam splitter with the $j$th and $k$th input beams is expressed in the form 
\begin{gather}
\label{BS}
    \hat A^\chi_{01}(\alpha,\theta)=\begin{pmatrix} 
                \cos{\chi} \,e^{i\theta} & \sin{\chi}\,e^{i(\theta+\alpha)} & 0 \\
                -\sin{\chi}\,e^{i(\theta-\alpha)} & \cos{\chi}\,e^{i\theta} & 0 \\
                0 & 0 & 1
            \end{pmatrix};\\
    \hat A^\chi_{12}(\alpha,\theta)=\begin{pmatrix} 
                1 & 0 & 0\\
                0 &  \cos{\chi}\,e^{i\theta} & \sin{\chi}\,e^{i(\theta+\alpha)} \\
                0 & -\sin{\chi}\,e^{i(\theta-\alpha)} &  \cos{\chi}\,e^{i\theta}
            \end{pmatrix},
\end{gather}
where $\chi$ determines the split ratio ($\sqrt{T}=\cos{\chi}$, $\sqrt{R}=\sin{\chi}$); $\alpha$ and $\theta$ are certain phases. 
The matrix $\hat P_j^\beta$ corresponding to the phase change by $\beta$ of the $j$-th beam is defined as
\begin{equation}
    \hat P_{0,1,2}^\beta=\Big\{\begin{pmatrix} 
                e^{i\beta} & 0 & 0\\
                0 & 1 & 0\\
                0 & 0 & 1
            \end{pmatrix};\,
            \begin{pmatrix} 
                1 & 0 & 0\\
                0 & e^{i\beta} & 0\\
                0 & 0 & 1
            \end{pmatrix};\,
            \begin{pmatrix} 
                1 & 0 & 0\\
                0 & 1 & 0\\
                0 & 0 & e^{i\beta}
            \end{pmatrix}\Big\}.
\end{equation}
In order to prepare a beam splitter matrix with an arbitrary desired ratio of reflection to transmission, one
has to assemble a Mach--Zehnder interferometer using two symmetric 50:50 beam splitters (for convenience, hereinafter we will omit the notation for dependence on $\alpha$ and $\theta$ if $(\alpha,\theta)=(\pi/2,0)$):
\begin{equation}
    \hat A^\chi_{01} \equiv \hat A^\chi_{01}(\pi/2,0)=\hat P_0^{\chi+\pi}\,\hat P_1^{\chi+3\pi/2}\,\hat A^{\pi/4}_{01}
    \,\hat P_0^{\pi-2\chi}\,\,\hat A^{\pi/4}_{01}\,\hat P_1^{\pi/2},
\end{equation}
with $\hat A^{\pi/4}_{01}$ corresponding to the ideal symmetric beam splitter.

As shown in Ref.\,[\citeonline{Suslov2011}], the Fourier transformation $\hat{F}$ can be factorized as follows:
\begin{align}
\label{susq3}
    \hat{F} = \hat P_1^{\pi/2}\,\hat A^{\pi/4}_{12}\,\hat P_0^\pi\,\hat A^{\tilde\chi}_{01}\,\hat P_0^\pi\,\hat P_1^{\pi/2}\,\hat A^{\pi/4}_{12}\,\hat P_2^{3\pi/2}
    =\hat P_0^{\tilde\chi+\pi}\,\hat P_1^{\pi/2}\,\hat A^{\pi/4}_{12}\,\hat P_1^{\pi/2+\tilde\chi}\,\hat A^{\pi/4}_{01}
    \,\hat P_0^{\pi-2\tilde\chi}\,\hat A^{\pi/4}_{01}\,\hat A^{\pi/4}_{12}\,\hat P_2^{3\pi/2};
\end{align}
where $\tilde\chi = \tan^{-1}\left(\sqrt{2}\right)$.
It is seen from this expression that the experimental realization of $\hat{F}$ requires no more than 4 symmetric 50:50 beam splitters. 
The optical circuit for $\hat{F}$ is depicted in Fig.\,\ref{qutrit_circ}.

\section*{Experimental setup}
\label{experiment}
The experimental layout is divided into two modules, as shown in Fig.\,\ref{scheme}.
In the state preparation module, the incident laser beam is converted into the qutrit initial state given by Eq.\,(\ref{initial_state}).
The beam splitters BS$^a$ and BS$^b$ generate three beams each representing a particular basis state $\ket{j}$ ($j=\{0,1,2\}$).
The $\ket{1}$ and $\ket{2}$ beams then pass through respectively one (PS$_\phi$) and two (PS$_{2\phi}$) phase shifters attached to a swivel platform which sets the relative phases $0$, $\phi$ and $2\phi$.
The value of $\phi$ depends on the position of the platform: by rotating the platform one alters the length of the optical paths through the phase shifters and, therefore, changes $\phi$ without affecting the ratio between the relative phases.

The primary module shown in Fig.\,\ref{scheme} realizes Eq.\,(\ref{susq3}).
However, although Eq.\,(\ref{susq3}) directly translates the Fourier transformation into the optical setting, it fails to take account of limitations intrinsic to the real equipment.
Namely, the transmission the phase shifters is associated with the intensity losses.
In order to take such losses into account we should employ the corresponding operators $\hat L^t_{0,1,2}$:
\begin{equation*}
\hat L_{0,1,2}^{t}=\Big\{\begin{pmatrix} 
t & 0 & 0\\
0 & 1 & 0\\
0 & 0 & 1
\end{pmatrix};\,
\begin{pmatrix} 
1 & 0 & 0\\
0 & t & 0\\
0 & 0 & 1
\end{pmatrix};\,
\begin{pmatrix} 
1 & 0 & 0\\
0 & 1 & 0\\
0 & 0 & t
\end{pmatrix}\Big\},
\end{equation*}
where $t$ is the absolute value of the transmission coefficient of and individual phase shifter.
After the appropriate alignments, the equation for the operation realized in the primary module assumes the form
\begin{equation}
\label{susq3_junk}
\hat{U}=
[\hat{A}_{12}^{\chi_0}(\alpha _{ 4 }, \theta_4)\, 
\hat{P}_2 ^ { \psi _6 } \,
\hat{P}_1 ^ { x _4 } \,
\hat{L}_1^{t_\text{ps}}\,
\hat{L}_2^{t_\text{ps}}\,
\hat{P}_1 ^ { \psi _5 } ]_4\,
[\hat{A}_{01}^{\chi_0}(\alpha _3, \theta_3) \,
\hat{P}_1 ^ { \psi _4 } \,
\hat{P}_0 ^ { x _3 } \,
\hat{L}_0^{t_\text{ps}}\,
\hat{L}_1^{t_\text{ps}}\,
\hat{P}_0 ^ { \psi _3 }]_3\,
[\hat{A}_{01}^{\chi_0}(\alpha _2, \theta_2)\, 
\hat{P}_1 ^ { x_2 } \,
\hat{L}_1^{t_\text{ps}}\,
\hat{P}_1 ^ { \psi _2 } ]_2\,
[\hat{A}_{12}^{\chi_0}(\alpha _1, \theta_1) \,
\hat{L}_ { 2 } ^{t_\text{ps}}\,
\hat{P}_2 ^ { x _1 } \,
\hat{P}_2 ^ { \psi _1 } \, 
]_1,
\end{equation}
where $t_\text{ps}$ is the modulus of the transmission coefficient of PS$_1$, \dots, PS$_4$; $\chi_0$ defines the beam splitters' split ratio ($\sqrt{T}=\cos{\chi_0}$, $\sqrt{R}=\sin{\chi_0}$); $\alpha_i$ and $\theta_i$ correspond to BS$^i$ (see Eq.\,(\ref{BS})); $\psi_i$ is the phase change due to reflection of M$_i$; $x_i$ is the phase change on PS$_i$.
In our experiment $t_\text{ps} = 0.935$, $T=0.445$ and $R=0.555$.
The notation $[\dots]_i$ will be used later.
For simplicity, the above formula does not explicitly include discrepancies in the optical distances.
In this respect, we should define $x_i$ as a relative phase in which such terms along with the phase shift on PS$_i$ are taken into account.
The output state vector can be written as
\begin{equation}
    \label{exp_out}
    \ket{\Psi_\text{out} (x_1, x_2, x_3, x_4, \phi)} = \hat{U}\{\hat{P}_{0}  ^ { \psi _{ a } }
    \hat{L}_ {1} ^ {t_{\phi}}
    \hat{P}_{1}  ^ { \phi } 
    \hat{L}_ {2} ^ {t_{ND}} 
    \hat{L}_ {2} ^ {t_{2\phi}}
    \hat{P}_ { 2 } ^ { 2 \phi }
    \hat{A}_{01}^{\chi_0}(\alpha _a, \theta_a)
    \hat{A}_{02}^{\chi_0}(\alpha _b, \theta_b)
    \ket{2}\}_\text{sp},
\end{equation}
where the brackets $\{\dots\}_\text{sp}$ denote the state prepared in the first module of the scheme;
$t_{\phi}$ are $t_{2\phi}$ are the absolute values of the transmission coefficient of PS$_\phi$ and PS$_{2\phi}$ respectively (in our experiment $t_{\phi} = 0.875$, $t_{2\phi} = 0.894$);
$t_{ND}=0.837$ is the modulus of the transmission coefficient of neutral-density (ND) filter, used for leveling of the intensities;
$(\alpha_a,\theta_a)$, $(\alpha_b,\theta_b)$ and $\psi_a$ correspond respectively to BS$^a$, BS$^b$ and M$_a$.

For certain values of $x_i$ which we denote by $x^F_i$ and which are given by
\begin{align}
    x^F_1=&-\alpha_1+\alpha_a-\alpha_b+\theta_b-\psi_1+\pi;\notag\\
    x^F_2=&-\alpha_2+\alpha_b-\theta_1+\psi_a-\psi_2-\pi/2;\notag\\
    x^F_3=&-\alpha_2+\alpha_3-\psi_3+\psi_4+\pi-2\tilde\chi;\notag\\
    x^F_4=&-\alpha_1+\alpha_2+\alpha_4-\alpha_b+\theta_1-\theta_2-\theta_3-\psi_4-\psi_5+\psi_6-\psi_a-\pi+\tilde\chi,
    \label{xF}
\end{align}
the transformation implemented in the scheme is similar to Eq.\,(\ref{susq3}):
\begin{align}
\label{susq3_fixed}
\hat F_{exp} = [\hat{A}^{\chi_0}_{12}\,\hat{P}_1^{\pi/2+\tilde\chi}\,\hat{L}_1^{t_\text{ps}}\,\hat{L}_2^{t_\text{ps}}]_4\, [\hat{A}^{\chi_0}_{01}\,\hat{P}_0^{\pi - 2\tilde\chi}\,\hat{L}_0^{t_\text{ps}}\,\hat{L}_1^{t_\text{ps}}]_3\,
 [\hat{A}^{\chi_0}_{01}\,\hat{L}_1^{t_\text{ps}}]_2\,
 [\hat{A}^{\chi_0}_{12}\,\hat{L}_2^{t_\text{ps}}\,\hat{P}_2^{3\pi/2}]_1.
\end{align}
Here we ignored the phases of the resulting beams incident on the detectors.
For the description of the alignment procedure see Methods and SI.
Figure\,\ref{signals_theory} displays the theoretical plots obtained using Eqs.\,(\ref{susq3}) (dashed lines) and (\ref{susq3_fixed}) (solid lines).
Both series of plots are almost identical. 
Note, that taking losses into account in Eq.\,(\ref{susq3_fixed}) results in smaller secondary peaks.
\begin{figure*}[t]
    \noindent\centering{
    \includegraphics[width=170mm]{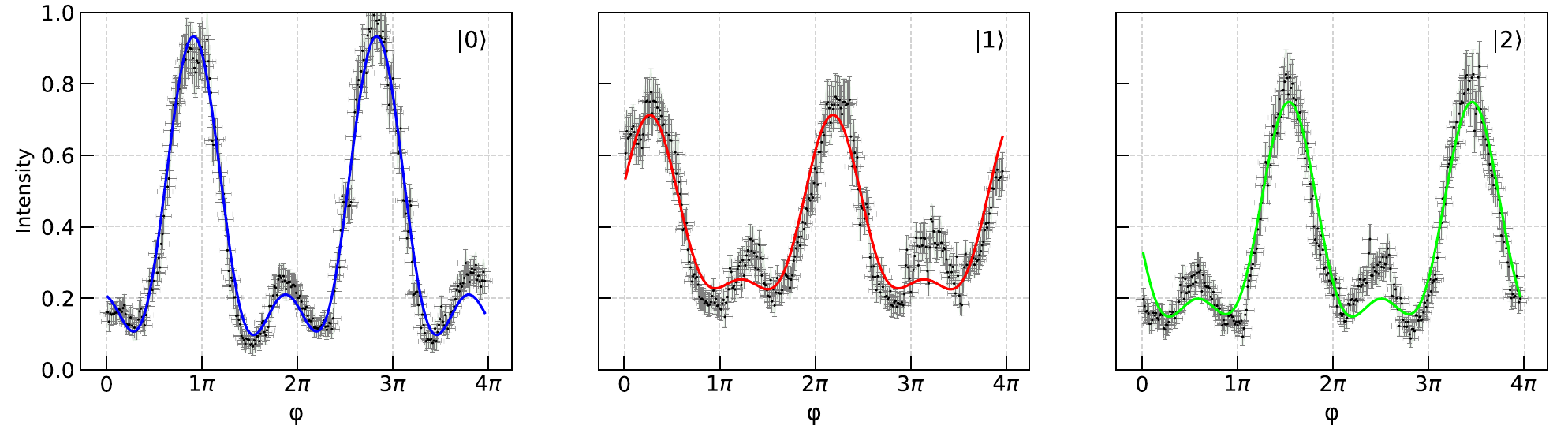}
    }
    \caption{
        The measured intensities on each of the detectors as functions of $\phi$. 
        The solid line shows the theoretical fit to the data.
        Each data point is calculated by averaging the experimental signal on the detector over $\sim 0.5$\,s with the fixed angle of the swivel platform; the vertical error bars represent the corresponding signal dispersion.
        The horizontal error bars reflect the precision limit of the swivel platform.
    }
    \label{results}
\end{figure*}

\section*{Results and Discussion}
\label{discussion}
Figure\,\ref{results} shows the measured intensities as functions of $\phi$.
The data on D$_i$ are fit by the square of $i$th element of the output vector function given by Eq.\,(\ref{exp_out}): 
\begin{equation}
\label{fit}
p_i = a_i \left| \braket{i|\Psi_\text{out} (x_1, x_2, x_3, x_4, \kappa \cdot \phi + \mu )} \right|^2 + b_i,
\end{equation}
where $a_i$ is the intensity scaling parameter; $b_i$ is the intensity bias simulating the interference visibility loss; $\kappa$ and $\mu$ are respectively the phase scaling parameter and phase shift independent of $i$.
The fitting is done using the method of least squares.
Note, that the same unitary transformation can be realized with the different sets of parameters.
For details on fitting and determining the corresponding errors see SI and Methods.
The phases $x_i$ determined from the fit are given by
\begin{equation}
    (x_1,x_2,x_3,x_4) = (x^F_1,x^F_2,x^F_3,x^F_4) \pm (0.28,  0.30,  0.30,  0.28),
    \label{fit_eq}
\end{equation}
where the second term is the error of fitting.
Despite the discrepancies (which, as a matter of fact, are small as compared to $\pi$) described by the second term, our data compare fairly well with the theoretical plots presented in Fig.\,\ref{signals_theory}.
The results show that the interference is controlled to the high degree in spite of the complexity of the optical scheme.
Thus, the described optical platform proves to be capable to perform small scale unitary operations.

Note that
our experiment is carried out in a multiphoton rather than in a single-photon regime which typically serves as a bedrock for the optical implementations of the quantum algorithms.
At the same time, similarly to many existing quantum algorithms, the realized Fourier phase-estimation protocol relies on the wave interference effects although it does not utilize specifically quantum phenomena.
As any lossless quantum computation, the Fourier transform is described by the unitary operator\,\cite{NielsenChuang}. We have constructed such a unitary operator through the specific arrangement of linear optical elements. 
It should be noted, however, that the discussed multiphoton approach does not support the algorithms relying on choosing between quantum alternatives (which takes place, for example, in the quantum random number generation procedure). 

As shown in Ref.\,[\citeonline{Reck}], the number of beam splitters needed to construct a general $N$-dimensional unitary matrix $U$ grows as $N(N-1)/2$.
The practical realization of such a multiport architecture, however, imposes additional scalability limitations (see SI for the detailed quantitative analysis):
\begin{enumerate}
    \item \textit{Restricted phase adjustment precision. }
    The relative cumulative error in the constructed matrix $U$ caused by the limited precision $\Delta\alpha$ with which we control the rotation angle of the optical holders and the width $d$ of the phase shifters is of order $N (d/\lambda)(n-1) \Delta \alpha$.
     Here $n$ is the refractive index of the phase shifers, and $\lambda$ is the light wavelength.
    \item \textit{Restricted precision of the wavefronts' alignment.}
    The misalignment of the wavefronts results in the complex interference pictures which can no longer be considered one-dimensional.
    The visibility of the picture deteriorates with the factor 
    $\approx 1 - \frac{N(R\Delta \alpha \frac{2\pi}{\lambda})^2}{8}$, where $R$ is the size of the beam spot.
    \item \textit{Phase fluctuations caused by the surface roughness.} 
    Assuming that the light acquires the delta-correlated random phase $\delta \xi$ due to the surface roughness of the optical elements, we estimate the corresponding visibility deterioration factor as $\sim e^{-N \langle \delta \xi^2 \rangle}$.
    \item \textit{Intensity losses.} The intensity losses on the mirrors and beam splitters used in our experiment are about $1\%$, which is acceptable.
    By far larger losses ($\approx 10\%$) are associated with the phase shifters.
    Nevertheless, the use of anti-reflective coating would reduce these losses to $1\%$.
    The signal intensity on the detector would be $0.99^N \approx \exp{\left( -N/100 \right)}$.
\end{enumerate}
Based on these estimates, a detailed quantitative analysis devises the prospect for realizing matrices with $N$ up to of order 100 (this upper limit is set mostly by item 4, for other details see SI).
This improvement will be built on the enhanced experimental and theoretical framework comprising the advanced adjustment precision of optical holders, eliminating the elements' surface roughness, minimizing intensity losses (e.g., via employing the anti-reflective coating), and mitigating the drift of phases caused by the mechanical oscillations and instability of the optical elements.
Note that the latter issue results in the rising deviation between the data and the fit as seen in Fig.\,\ref{signals_theory}. The corresponding improvement will be achieved by implementing the mechanical feedback phase control.

Further refining the concert between the theoretical description and the experimental realization will be achieved via including into the scheme the machine learning algorithms capable to compensate the imprecision in the alignment of the optical elements. 
These techniques have already passed the reliability test in the base-4 (ququart) version of the setup which we have already successfully realized.
The obtained results manifest the improved accuracy and serve as the evidence of the scheme's scalability.
The detailed description of the ququart experiment will be the subject of the forthcoming publication.

\begin{figure*}[t]
    \noindent\centering{
    \includegraphics[width=150mm]{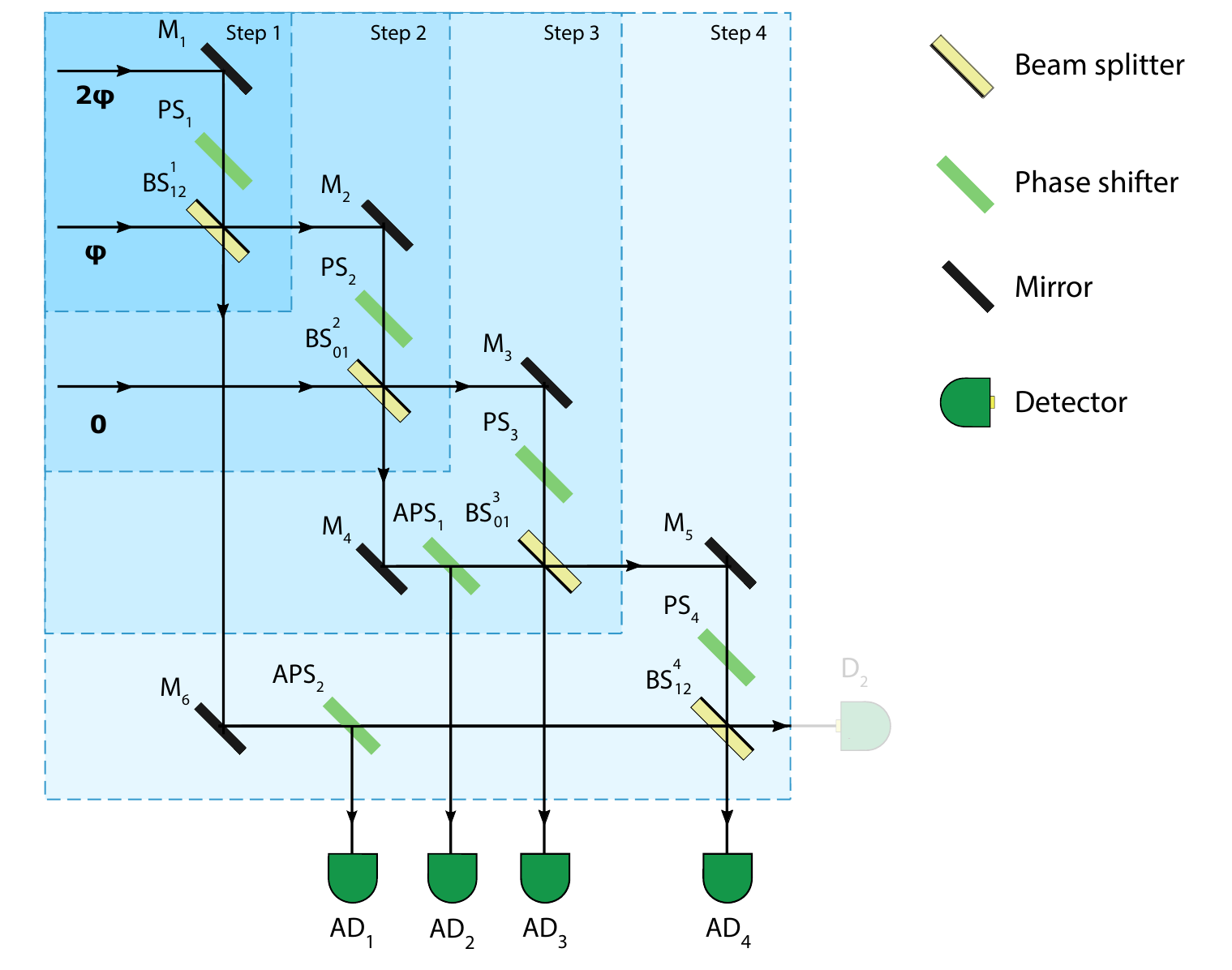}
    }
    \caption{
    Four consecutive steps of the alignment procedure. 
    At $i$th step, the output signal from the corresponding sector of the scheme (measured by the detector $AD_i$) is tuned to comply with the theoretical value calculated through the breakdown of Eq.\,(\ref{susq3_fixed}).
    The tuning is done through the alignment of $PS_i$.
    }
\label{method}
\end{figure*}

There have been a recent progress in demonstrating the advantage of Quantum Fourier transforms (QFT) interferometers using both path and polarization modes\,\cite{Su2017} and in realizing interferometric phase estimation algorithm approaching the Heisenberg limit\,\cite{Daryanoosh}. Our scheme employs larger number of linear optic elements as compared to above references and utilizing path modes only. Yet we achieved a fairly high level of the correspondence between the experiment and theory. By adding the polarization degrees of freedom analogously to\,\cite{Su2017,Daryanoosh}, we will further increase the dimension of the unitary matrix realized by our scheme.

\section*{Methods}

	Our optical setup includes the following equipment:
	\\
	\textit{Phase shifters.--} The phase shifters mainly serve to adjust the relative phases of the beams.
	In our setup, we use pieces of thick glass; the intensity loss on these elements is near $12.5\%$.
\\
	\textit{Beam splitters.--} We employ beam splitters with dielectric coating optimized for the $400 - 700\,$nm range.
	The nominal split ratio is 50:50.
	In practice however, this holds only if the incident laser beam is unpolarized. 
	For the case of the linearly polarized beam used in our experiment, the split ratio is close to 55:45.
	\\
	\textit{Mirrors.--} Dielectric mirrors optimized for the $400 - 700\,$nm range.
	\\
	\textit{Laser.--} Diode pumped solid state laser, 532\,nm, 150\,mW; the coherence length of light is 50\,m.
	\\
	\textit{Detectors.--} Photodiode detectors. 
\\
The use of the photodiodes for the detection is justified by their high measurement speed (as opposed to the single-photon detectors), as the main goal of the present work was to test the interference capacity of the complex optical setup.
However, an additional testing series employing the single-photon detectors has shown practically the same results as presented.
This fact means that such a replacement does not pose any significant changes in the operation of the circuit.

The alignment of the scheme is done in accordance with a step-by-step procedure which lays in tuning the signal at the intermediate points of the beams' paths (see Fig.\,\ref{method}).
At each consecutive step, the interference intensity at the given point is matched with the theoretical value obtained through the breakdown of Eq.\,(\ref{susq3_fixed}): the $i$th step of the procedure leverages the $i$th block of operators ($[\dots]_i$) in the relation.
At the first two stages, we receive the signal reflected from the phase shifters PS$_2$ and PS$_3$ using the detectors AD$_1$ and AD$_2$, respectively.
In turn, the last two stages involve the signals from the detectors AD$_3$ and AD$_4$.
The alignment is performed via rotating the phase shifters (i.e., altering the optical path length) preceding the given point.
By doing so, one changes the phases $x_i$ which in the end should be equal to $x_i^F$ given by Eq.\,(\ref{xF}).
For details see SI.

Each point in Fig.\,\ref{results} is obtained by averaging signals from the detectors generated over $\sim 0.5$\,s.
The oscillations and instability of the optical elements are represented by the vertical error bars. 
To estimate the corresponding error, we measured the signal discrepancies appearing over a characteristic period of time ($\sim 0.5$\,s) with the fixed angle of the swivel platform (determining the value of $\phi$).
The horizontal error bars express the limited precision of the swivel platform.

The fitting of the experimental data is done via applying the method of least squares, see more detail in SI.
The phases $\mathbf{x}=[x_1,x_2,x_3,x_4]^T$ corresponding to the optimal fit turned out to be very close to $\mathbf{x}^F=[x_1^F,\,x_2^F,\,x_3^F,\,x_4^F]^T$.
The fitting error of $x_k$ ($k\in\{1,\,2,\,3,\,4\}$) is determined by the maximum size of the neighbourhood $\mathcal{O}^F_k$ of $x^F_k$ such that for any $\tilde{x}_k \in \mathcal{O}^F_k$ the standard deviation of $p_i(\tilde{\mathbf{x}},\,\phi)$ (with $\tilde{\mathbf{x}}=[x_1^F,\,\dots, \tilde{x}_k^F,\dots, x_4^F]^T$) from $p_i(\mathbf{x}^F,\,\phi)$ ($i\in\{0,\,1,\,2\}$) does not exceed the experimental error.


\section*{Acknowledgements}
We thank Andrey Elagin, Sandy Heinz and Scott Wakely for furnishing facilities at the UChicago Enrico Fermi Institute where part of this work was completed.

This work was supported by the Government of the Russian Federation
(Agreement 05.Y09.21.0018),
by the RFBR Grants No. 17-02-00002 (M.V.L. and O.V.M.), 
17-02-00396A, 18-02-00642A and 19-32-80005 (N.S.K. and M.R.P.),
Foundation for the Advancement of Theoretical Physics and Mathematics "BASIS",
the Ministry of Education and Science of the Russian Federation
16.7162.2017/8.9, and by NSF grant DMR1809188 (N.S.K.).
The work of V.M.V. was supported by the U.S. Department
of Energy, Office of Science, Basic Energy Sciences, Materials Sciences and Engineering Division.

\section*{Author contributions statement}
V.V.Z., N.S.K., M.R.P., O.V.M., M.V.L. and G.B.L. conceived and planned the research. 
V.V.Z., O.V.M, M.V.L. and G.B.L. carried out the main part of the fieldwork. 
V.V.Z., N.S.K., D.I.L., V.M.V. and G.B.L. analyzed data and discussed results.
N.S.K. and V.M.V. wrote the manuscript.
All authors reviewed the manuscript.

\section*{Additional information}
The authors declare no competing interests.

\section*{Data availability statement}
All data generated or analyzed during this study are included in this published article and its Supplementary Information file.

\clearpage
\section*{Supplementary Information}
\setcounter{figure}{0} 
\setcounter{equation}{0}
\renewcommand{\theequation}{S\arabic{equation}}
\renewcommand{\thefigure}{S\arabic{figure}}
\bigskip
\subsection*{Alignment}
\label{ap1}
In this section we describe the alignment procedure for the qutrit quantum Fourier transformation setup. 
Our step-by-step approach lays in tuning the signal at the intermediate points of the beams' paths (see Fig.\,\ref{method} in the main text).
At each consecutive step, the interference intensity at the given point is matched with the theoretical value obtained through the breakdown of Eq.\,(13) from the main text.
At the first two stages, we receive the signal reflected from the phase shifters APS$_2$ and APS$_1$ using the detectors AD$_1$ and AD$_2$, respectively.
In turn, the last two stages involve the signals from the detectors AD$_3$ and AD$_4$.
The alignment is performed via rotating the phase shifters (i.e., altering the optical path length) preceding the given point.
By doing so, one changes the phases $x_i$ which in the end should be equal to $x_i^F$ given by Eq.\,(12) from the main text.

Let us now examine each step of the procedure in details.
\begin{figure*}[t]
    \noindent\centering{
    \includegraphics[width=150mm]{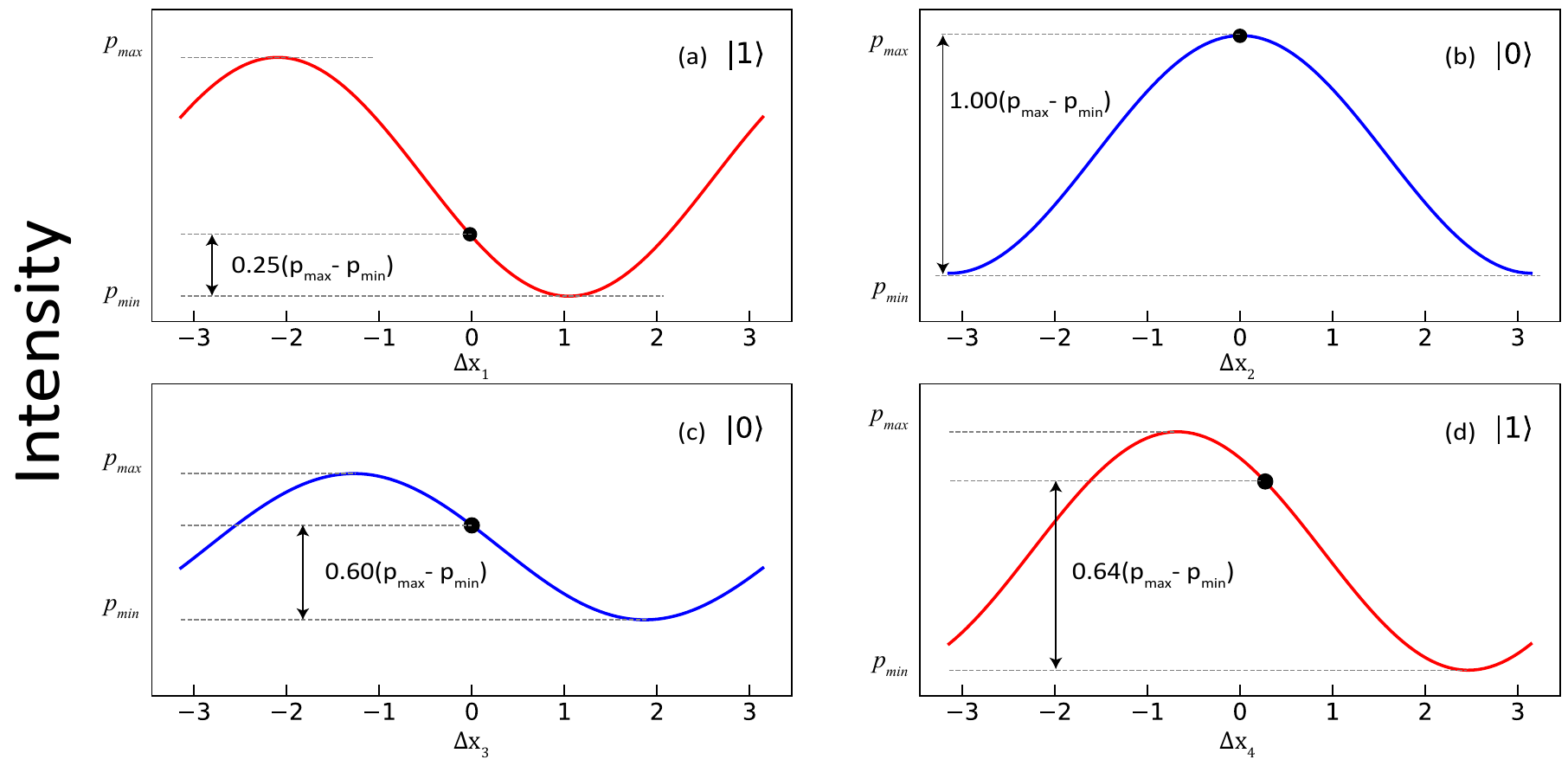}
    }
    \caption{
        Alignment plots of intensities on various detectors as functions of $\Delta x_i$ ($i=\{1,2,3,4\}$); $\Delta x_i$ is determined by the position of PS$_i$. 
        Black points correspond to the Fourier transform configuration.
        (a) Detector AD$_1$; first step of alignment. 
        (b) Detector AD$_2$; second step.
        (c) Detector AD$_3$; third step.
        (d) Detector AD$_4$; fourth step.
    }
    \label{adj}
\end{figure*}
\\
\textit{Step 1.--} 
Since $x_1$, $x_2$, $x_3$ and $x_4$ essentially determine the initial relative phases between the $\ket{0}$, $\ket{1}$ and $\ket{2}$ beams, we have a freedom in choosing $\phi$.
This is easily seen by noticing from  Eq.\,(11) of the main text that
\begin{equation}
\label{arbitrary_phase}
    p_i ( \phi, \Delta x_1, \Delta x_2, \Delta x_3, \Delta x_4 ) = p_i ( 0, \Delta x_1 + \phi, \Delta x_2 + \phi, \Delta x_3, \Delta x_4 - \phi ),
\end{equation}
with $i=\{0,1,2\}$;
here we denote $\Delta x_j = x_j-x^F_j$ ($j=\{1,2,3,4\}$).
Although the experimental value of $\phi$ (which shall be denoted $\phi^{real}$) set by $PS_{\phi}$ and $PS_{2\phi}$ is unknown, we can reassign $\phi$ to an arbitrary value. 
According to Eq.\,(\ref{arbitrary_phase}), in order to put $\phi=\phi_0$, we should renominate the target phases $x^F_i$ as follows:
\begin{align}
    &\widetilde{x}^F_1=x^F_1-\phi^{real}+\phi_0;\notag\\
    &\widetilde{x}^F_2=x^F_2-\phi^{real}+\phi_0;\notag\\
    &\widetilde{x}^F_3=x^F_3;\notag\\
    &\widetilde{x}^F_4=x^F_4+\phi^{real}-\phi_0.
    \label{xF2}
\end{align}
Here and throughout the whole procedure we put $\phi=\pi/3$.

Using AD$_1$ we measure the intensity of the $\ket{1}$ beam after it passes through BS$^1_{12}$.
This intensity may be regarded as the probability $\tilde{p}_1$ of finding the qutrit in the state $\ket{1}$ after the action of the first block of operators (denoted by $[\dots]_1$) in Eq.\,(10) and may be written
\begin{equation}
\label{p1}
\tilde{p}_1 = \sin ^2(\chi_0 ) \cos ^2(\chi_0 ) (1 - t_\text{ps} t_{2\phi} t_\text{f} (t_\text{ps} t_{2\phi} t_\text{f} + 2 t_{\phi} \sin (\chi_0 ) \cos (\Delta x_1+\phi ))-t_{\phi}^2 \sin ^2(\chi_0 )),
\end{equation}
Our object is to set the value of $\Delta x_1$ to zero so that the measured signal would comply with the action of the first block in Eq.\,(13).
Experimentally we achieve this by rotating PS$_1$ and controlling the intensity on AD$_1$.
According to Eq.\,(\ref{p1}), the target intensity can be expressed in terms of the experimentally measurable values as $\tilde{p}_1 = \min_{x_1} \tilde{p}_1 + 0.25\,(\max_{x_1} \tilde{p}_1-\min_{x_1} \tilde{p}_1)$.
Fig.\,\ref{adj}(a) shows the theoretical plot of the signal as function of $\Delta x_1$, where the dot marks the point to which we adjust PS$_1$.
\\
\textit{Step 2.--} Using AD$_2$ we measure the intensity of the $\ket{0}$ beam after it passes through BS$^2_{01}$.
Bearing in mind the second block of operators ($[\dots]_2$), we write the corresponding probability $\tilde{p}_2$:

\begin{multline}
\tilde{p}_2=
\frac{1}{32} (\sin ^2(2 \chi _0) (8 t_{\text{ps}} \sin (2 \chi _0) (t_f t_{\text{ps}} t_{2 \phi } (\sin
   (\text{$\Delta $x}_2+2 \phi )+t_{\text{ps}} t_{\phi } \cos (\chi _0) \cos (\phi ))\\+ \sin (\text{$\Delta $x}_2)
   t_{\phi } \sin (\chi _0) \cos (\phi ))+4 \cos (2 \chi _0) (t_f^2 t_{\text{ps}}^4 t_{2 \phi }^2-1)\\+4 t_f^2
   t_{\text{ps}}^4 t_{2 \phi }^2-t_{\text{ps}}^2 t_{\phi }^2 \cos (4 \chi _0)+t_{\text{ps}}^2 t_{\phi }^2+4)+64 \cos
   (\text{$\Delta $x}_2) t_{\text{ps}} t_{\phi } \sin ^4(\chi _0) \cos ^3(\chi _0) \sin (\phi ))
\end{multline}

The condition $\Delta x_2 = 0$ corresponds to a maximum of $\tilde{p}_2$ (see Fig.\,\ref{adj}(b)).
\\
\textit{Step 3.--} Using AD$_3$ we measure the intensity of the $\ket{0}$ beam after it passes through BS$^3_{01}$.
The corresponding probability $\tilde{p}_3$ after the action of the third block operators ($[\dots]_3$) is given by
\begin{multline}
    \tilde{p}_3 = 
    \frac{1}{32} (\frac{32}{3} t_{\text{ps}}^2 t_{\phi } \sin ^4(\chi _0) \cos ^3(\chi _0) \cos (\phi ) (2 (\sin
  (\text{$\Delta $x}_3)+2 \sqrt{2} \cos (\text{$\Delta $x}_3))\\
  +t_f t_{\text{ps}} t_{2 \phi } \sin (2 \chi _0)
  (-4 \sqrt{2} \sin (\text{$\Delta $x}_3) t_{\text{ps}}+2 \cos (\text{$\Delta $x}_3) t_{\text{ps}}+3
  t_{\text{ps}}^2+3))\\
  +\frac{1}{4} (\frac{8}{3} t_{\text{ps}} \sin ^4(2 \chi _0) (\cos (\text{$\Delta
  $x}_3)-2 \sqrt{2} \sin (\text{$\Delta $x}_3)) (2 t_f^2 t_{\text{ps}}^4 t_{2 \phi }^2-t_{\text{ps}}^2 t_{\phi }^2 \cos
  (2 \chi _0)+t_{\text{ps}}^2 t_{\phi }^2-2)\\
  +\frac{1}{4} (2 (t_{\text{ps}}^2+1) \cos (8 \chi _0) (2
  t_f^2 t_{\text{ps}}^4 t_{2 \phi }^2+t_{\text{ps}}^2 t_{\phi }^2-2)-8 (t_{\text{ps}}^2+1) \cos (4 \chi _0) (2 t_f^2
  t_{\text{ps}}^4 t_{2 \phi }^2+t_{\text{ps}}^2 t_{\phi }^2+2)\\
  +2 (t_{\text{ps}}^2+1) (6 t_f^2 t_{\text{ps}}^4 t_{2 \phi }^2+3
  t_{\text{ps}}^2 t_{\phi }^2+10)+t_{\text{ps}}^2 (t_{\text{ps}}^2+1) t_{\phi }^2 (-\cos (10 \chi _0))-2 \cos
  (2 \chi _0) (t_{\text{ps}}^2 ((t_{\text{ps}}^2+1) t_{\phi }^2-8)+8)\\
  +\cos (6 \chi _0)
  (t_{\text{ps}}^2 (3 (t_{\text{ps}}^2+1) t_{\phi }^2-16)+16)))\\
  -4 t_f t_{\text{ps}}^2 t_{2 \phi } \sin
  ^3(2 \chi _0) \sin (2 \phi ) (\cos (2 \chi _0) (\frac{2}{3} t_{\text{ps}} (\cos (\text{$\Delta
  $x}_3)-2 \sqrt{2} \sin (\text{$\Delta $x}_3))+t_{\text{ps}}^2+1)+t_{\text{ps}}^2-1)\\
  +8 t_f t_{\text{ps}}^3 t_{2
  \phi } \sin ^3(2 \chi _0) \cos (2 \phi ) \sin (2 \tan ^{-1}(\sqrt{2})-\text{$\Delta $x}_3)\\
  -32 t_{\text{ps}} t_{\phi
  } \sin ^4(\chi _0) \cos ^3(\chi _0) \sin (\phi ) (\cos (2 \chi _0) (\frac{2}{3} t_{\text{ps}} (\cos
  (\text{$\Delta $x}_3)-2 \sqrt{2} \sin (\text{$\Delta $x}_3))+t_{\text{ps}}^2+1)+t_{\text{ps}}^2-1))
\end{multline}
For $\Delta x_3 = 0$ we have $\tilde{p}_3 = \min_{x_3} \tilde{p}_3 + 0.60\,(\max_{x_3} \tilde{p}_3-\min_{x_3} \tilde{p}_3)$ (see Fig.\,\ref{adj}(c)).
\\
\textit{Step 4.--} Using AD$_3$ we measure the intensity of the $\ket{1}$ beam after it passes through BS$^4_{12}$.
The corresponding probability $\tilde{p}_4$ after the action of the fourth block operators ($[\dots]_4$) is given by

\begingroup
\allowdisplaybreaks
\begin{multline}
    \tilde{p}_4 = \frac{1}{2048}(16 \cos (8 \chi _0) \sin ^2(\chi _0) t_f^2 t_{2 \phi }^2 t_{\text{ps}}^8+48 \sin ^2(\chi _0) t_f^2 t_{2 \phi }^2
   t_{\text{ps}}^8-16 \cos (2 \tan ^{-1}(\sqrt{2})-8 \chi _0) \sin ^2(\chi _0) t_f^2 t_{2 \phi }^2 t_{\text{ps}}^7
   \\
   -64
   \cos (2 \tan ^{-1}(\sqrt{2})-6 \chi _0) \sin ^2(\chi _0) t_f^2 t_{2 \phi }^2 t_{\text{ps}}^7-64 \cos (2 \tan
   ^{-1}(\sqrt{2})-4 \chi _0) \sin ^2(\chi _0) t_f^2 t_{2 \phi }^2 t_{\text{ps}}^7
   \\
   +64 \cos (2 \tan
   ^{-1}(\sqrt{2})-2 \chi _0) \sin ^2(\chi _0) t_f^2 t_{2 \phi }^2 t_{\text{ps}}^7+64 \cos (2 (\chi _0+\tan
   ^{-1}(\sqrt{2}))) \sin ^2(\chi _0) t_f^2 t_{2 \phi }^2 t_{\text{ps}}^7
   \\
   -64 \cos (4 \chi _0+2 \tan
   ^{-1}(\sqrt{2})) \sin ^2(\chi _0) t_f^2 t_{2 \phi }^2 t_{\text{ps}}^7-64 \cos (6 \chi _0+2 \tan
   ^{-1}(\sqrt{2})) \sin ^2(\chi _0) t_f^2 t_{2 \phi }^2 t_{\text{ps}}^7
   \\
   -16 \cos (8 \chi _0+2 \tan
   ^{-1}(\sqrt{2})) \sin ^2(\chi _0) t_f^2 t_{2 \phi }^2 t_{\text{ps}}^7-\frac{160}{3} \sin ^2(\chi _0) t_f^2
   t_{2 \phi }^2 t_{\text{ps}}^7
   \\
   +48 \cos (\phi ) \sin (\chi _0) t_f t_{\phi } t_{2 \phi } t_{\text{ps}}^7-4 \cos (\phi -10 \chi
   _0) \sin (\chi _0) t_f t_{\phi } t_{2 \phi } t_{\text{ps}}^7+8 \cos (\phi -8 \chi _0) \sin (\chi _0) t_f
   t_{\phi } t_{2 \phi } t_{\text{ps}}^7
   \\
   +12 \cos (\phi -6 \chi _0) \sin (\chi _0) t_f t_{\phi } t_{2 \phi } t_{\text{ps}}^7-32
   \cos (\phi -4 \chi _0) \sin (\chi _0) t_f t_{\phi } t_{2 \phi } t_{\text{ps}}^7-8 \cos (\phi -2 \chi _0) \sin
   (\chi _0) t_f t_{\phi } t_{2 \phi } t_{\text{ps}}^7
   \\
   -8 \cos (\phi +2 \chi _0) \sin (\chi _0) t_f t_{\phi } t_{2 \phi
   } t_{\text{ps}}^7-32 \cos (\phi +4 \chi _0) \sin (\chi _0) t_f t_{\phi } t_{2 \phi } t_{\text{ps}}^7+12 \cos (\phi +6
   \chi _0) \sin (\chi _0) t_f t_{\phi } t_{2 \phi } t_{\text{ps}}^7
   \\
   +8 \cos (\phi +8 \chi _0) \sin (\chi _0) t_f
   t_{\phi } t_{2 \phi } t_{\text{ps}}^7-4 \cos (\phi +10 \chi _0) \sin (\chi _0) t_f t_{\phi } t_{2 \phi } t_{\text{ps}}^7+12
   \cos (6 \chi _0) t_{\phi }^2 t_{\text{ps}}^6+2 \cos (8 \chi _0) t_{\phi }^2 t_{\text{ps}}^6
   \\
   -4 \cos (10 \chi _0)
   t_{\phi }^2 t_{\text{ps}}^6+\cos (12 \chi _0) t_{\phi }^2 t_{\text{ps}}^6+128 \cos (6 \chi _0) \sin ^2(\chi _0)
   t_f^2 t_{2 \phi }^2 t_{\text{ps}}^6+16 \cos (8 \chi _0) \sin ^2(\chi _0) t_f^2 t_{2 \phi }^2 t_{\text{ps}}^6
   \\
   +560 \sin
   ^2(\chi _0) t_f^2 t_{2 \phi }^2 t_{\text{ps}}^6-512 \cos (\phi ) \cos (\chi _0) \sin (\phi ) \sin ^3(\chi _0) t_f
   t_{2 \phi } t_{\text{ps}}^6+64 \cos (\chi _0) \sin (2 \phi -6 \chi _0) \sin ^3(\chi _0) t_f t_{2 \phi }
   t_{\text{ps}}^6
   \\
   +128 \cos (\chi _0) \sin (2 \phi -4 \chi _0) \sin ^3(\chi _0) t_f t_{2 \phi } t_{\text{ps}}^6-64
   \cos (\chi _0) \sin (2 \phi -2 \chi _0) \sin ^3(\chi _0) t_f t_{2 \phi } t_{\text{ps}}^6
   \\
   -64 \cos (\chi
   _0) \sin ^3(\chi _0) \sin (2 (\phi +\chi _0)) t_f t_{2 \phi } t_{\text{ps}}^6+128 \cos (\chi _0)
   \sin ^3(\chi _0) \sin (2 \phi +4 \chi _0) t_f t_{2 \phi } t_{\text{ps}}^6
   \\
   +64 \cos (\chi _0) \sin ^3(\chi
   _0) \sin (2 \phi +6 \chi _0) t_f t_{2 \phi } t_{\text{ps}}^6+48 \cos (\phi -2 \tan ^{-1}(\sqrt{2})) \sin
   (\chi _0) t_f t_{\phi } t_{2 \phi } t_{\text{ps}}^6
   \\
   +48 \cos (\phi +2 \tan ^{-1}(\sqrt{2})) \sin (\chi
   _0) t_f t_{\phi } t_{2 \phi } t_{\text{ps}}^6+4 \cos (\phi -10 \chi _0+2 \tan ^{-1}(\sqrt{2})) \sin (\chi _0)
   t_f t_{\phi } t_{2 \phi } t_{\text{ps}}^6
   \\
   +8 \cos (\phi -8 \chi _0-2 \tan ^{-1}(\sqrt{2})) \sin (\chi _0) t_f
   t_{\phi } t_{2 \phi } t_{\text{ps}}^6+8 \cos (\phi -8 \chi _0+2 \tan ^{-1}(\sqrt{2})) \sin (\chi _0) t_f t_{\phi }
   t_{2 \phi } t_{\text{ps}}^6
   \\
   -12 \cos (\phi -6 \chi _0-2 \tan ^{-1}(\sqrt{2})) \sin (\chi _0) t_f t_{\phi } t_{2 \phi
   } t_{\text{ps}}^6-12 \cos (\phi -6 \chi _0+2 \tan ^{-1}(\sqrt{2})) \sin (\chi _0) t_f t_{\phi } t_{2 \phi }
   t_{\text{ps}}^6
   \\
   -32 \cos (\phi -4 \chi _0-2 \tan ^{-1}(\sqrt{2})) \sin (\chi _0) t_f t_{\phi } t_{2 \phi }
   t_{\text{ps}}^6-32 \cos (\phi -4 \chi _0+2 \tan ^{-1}(\sqrt{2})) \sin (\chi _0) t_f t_{\phi } t_{2 \phi }
   t_{\text{ps}}^6
   \\
   +8 \cos (\phi -2 \chi _0+2 \tan ^{-1}(\sqrt{2})) \sin (\chi _0) t_f t_{\phi } t_{2 \phi }
   t_{\text{ps}}^6+8 \cos (\phi +2 \chi _0-2 \tan ^{-1}(\sqrt{2})) \sin (\chi _0) t_f t_{\phi } t_{2 \phi }
   t_{\text{ps}}^6
   \\
   -32 \cos (\phi +4 \chi _0-2 \tan ^{-1}(\sqrt{2})) \sin (\chi _0) t_f t_{\phi } t_{2 \phi }
   t_{\text{ps}}^6-32 \cos (\phi +4 \chi _0+2 \tan ^{-1}(\sqrt{2})) \sin (\chi _0) t_f t_{\phi } t_{2 \phi }
   t_{\text{ps}}^6
   \\
   -12 \cos (\phi +6 \chi _0-2 \tan ^{-1}(\sqrt{2})) \sin (\chi _0) t_f t_{\phi } t_{2 \phi }
   t_{\text{ps}}^6-12 \cos (\phi +6 \chi _0+2 \tan ^{-1}(\sqrt{2})) \sin (\chi _0) t_f t_{\phi } t_{2 \phi }
   t_{\text{ps}}^6
   \\
   +8 \cos (\phi +8 \chi _0-2 \tan ^{-1}(\sqrt{2})) \sin (\chi _0) t_f t_{\phi } t_{2 \phi }
   t_{\text{ps}}^6+8 \cos (\phi +8 \chi _0+2 \tan ^{-1}(\sqrt{2})) \sin (\chi _0) t_f t_{\phi } t_{2 \phi }
   t_{\text{ps}}^6
   \\
   +4 \cos (\phi +10 \chi _0-2 \tan ^{-1}(\sqrt{2})) \sin (\chi _0) t_f t_{\phi } t_{2 \phi }
   t_{\text{ps}}^6+4 \cos (\phi +10 \chi _0+2 \tan ^{-1}(\sqrt{2})) \sin (\chi _0) t_f t_{\phi } t_{2 \phi }
   t_{\text{ps}}^6
   \\
   +8 \cos (\phi -2 (\chi _0+\tan ^{-1}(\sqrt{2}))) \sin (\chi _0) t_f t_{\phi } t_{2 \phi }
   t_{\text{ps}}^6+8 \cos (\phi +2 (\chi _0+\tan ^{-1}(\sqrt{2}))) \sin (\chi _0) t_f t_{\phi } t_{2 \phi }
   t_{\text{ps}}^6
   \\
   +4 \cos (\phi -2 (5 \chi _0+\tan ^{-1}(\sqrt{2}))) \sin (\chi _0) t_f t_{\phi } t_{2 \phi
   } t_{\text{ps}}^6-\cos (2 \tan ^{-1}(\sqrt{2})-12 \chi _0) t_{\phi }^2 t_{\text{ps}}^5
   \\
   +6 \cos (2 \tan
   ^{-1}(\sqrt{2})-8 \chi _0) t_{\phi }^2 t_{\text{ps}}^5-15 \cos (2 \tan ^{-1}(\sqrt{2})-4 \chi _0) t_{\phi }^2
   t_{\text{ps}}^5-15 \cos (4 \chi _0+2 \tan ^{-1}(\sqrt{2})) t_{\phi }^2 t_{\text{ps}}^5
   \\
   -\cos (2 (6 \chi _0+\tan
   ^{-1}(\sqrt{2}))) t_{\phi }^2 t_{\text{ps}}^5+6 \cos (8 \chi _0+2 \tan ^{-1}(\sqrt{2})) t_{\phi }^2
   t_{\text{ps}}^5+256 \sin (\tan ^{-1}(\sqrt{2})-\text{$\Delta $x}_4) \sin ^2(\chi _0) t_f^2 t_{2 \phi }^2
   t_{\text{ps}}^5
   \\
   -64 \sin (-\text{$\Delta $x}_4-6 \chi _0+\tan ^{-1}(\sqrt{2})) \sin ^2(\chi _0) t_f^2 t_{2 \phi }^2
   t_{\text{ps}}^5-128 \sin (-\text{$\Delta $x}_4-4 \chi _0+\tan ^{-1}(\sqrt{2})) \sin ^2(\chi _0) t_f^2 t_{2 \phi }^2
   t_{\text{ps}}^5
   \\
   +64 \sin (-\text{$\Delta $x}_4-2 \chi _0+\tan ^{-1}(\sqrt{2})) \sin ^2(\chi _0) t_f^2 t_{2 \phi }^2
   t_{\text{ps}}^5+64 \sin ^2(\chi _0) \sin (-\text{$\Delta $x}_4+2 \chi _0+\tan ^{-1}(\sqrt{2})) t_f^2 t_{2 \phi }^2
   t_{\text{ps}}^5
   \\
   -128 \sin ^2(\chi _0) \sin (-\text{$\Delta $x}_4+4 \chi _0+\tan ^{-1}(\sqrt{2})) t_f^2 t_{2 \phi }^2
   t_{\text{ps}}^5-64 \sin ^2(\chi _0) \sin (-\text{$\Delta $x}_4+6 \chi _0+\tan ^{-1}(\sqrt{2})) t_f^2 t_{2 \phi }^2
   t_{\text{ps}}^5
   \\
   -96 \cos (\chi _0) \sin (\phi ) \sin ^2(\chi _0) t_{\phi } t_{\text{ps}}^5+64 \cos (\chi _0) \sin
   (\phi -4 \chi _0) \sin ^2(\chi _0) t_{\phi } t_{\text{ps}}^5-8 \sin (\phi -8 \chi _0) \sin (\chi _0)
   \sin (2 \chi _0) t_{\phi } t_{\text{ps}}^5
   \\
   +64 \cos (\chi _0) \sin ^2(\chi _0) \sin (\phi +4 \chi _0)
   t_{\phi } t_{\text{ps}}^5-8 \sin (\chi _0) \sin (2 \chi _0) \sin (\phi +8 \chi _0) t_{\phi } t_{\text{ps}}^5
   \\
   +256
   \cos (\chi _0) \sin (2 \phi -2 \tan ^{-1}(\sqrt{2})) \sin ^3(\chi _0) t_f t_{2 \phi }
   t_{\text{ps}}^5-1280 \cos (\chi _0) \sin (2 (\phi +\tan ^{-1}(\sqrt{2}))) \sin ^3(\chi _0)
   t_f t_{2 \phi } t_{\text{ps}}^5
   \\
   -64 \cos (\chi _0) \sin (2 \phi -6 \chi _0-2 \tan ^{-1}(\sqrt{2})) \sin ^3(\chi
   _0) t_f t_{2 \phi } t_{\text{ps}}^5
   \\
   -128 \cos (\chi _0) \sin (2 \phi -4 \chi _0-2 \tan ^{-1}(\sqrt{2})) \sin
   ^3(\chi _0) t_f t_{2 \phi } t_{\text{ps}}^5
   \\
   -64 \cos (\chi _0) \sin (2 (\phi -3 \chi _0+\tan
   ^{-1}(\sqrt{2}))) \sin ^3(\chi _0) t_f t_{2 \phi } t_{\text{ps}}^5
   \\
   +64 \cos (\chi _0) \sin (2 \phi
   -2 \chi _0-2 \tan ^{-1}(\sqrt{2})) \sin ^3(\chi _0) t_f t_{2 \phi } t_{\text{ps}}^5
   \\
   -384 \cos (\chi _0) \sin
   (2 (\phi -2 \chi _0+\tan ^{-1}(\sqrt{2}))) \sin ^3(\chi _0) t_f t_{2 \phi } t_{\text{ps}}^5
   \\
   -960 \cos
   (\chi _0) \sin (2 (\phi -\chi _0+\tan ^{-1}(\sqrt{2}))) \sin ^3(\chi _0) t_f t_{2 \phi }
   t_{\text{ps}}^5
   \\
   +64 \cos (\chi _0) \sin ^3(\chi _0) \sin (2 (\phi +\chi _0-\tan
   ^{-1}(\sqrt{2}))) t_f t_{2 \phi } t_{\text{ps}}^5
   \\
   -960 \cos (\chi _0) \sin ^3(\chi _0) \sin (2
   (\phi +\chi _0+\tan ^{-1}(\sqrt{2}))) t_f t_{2 \phi } t_{\text{ps}}^5
   \\
   -384 \cos (\chi _0) \sin ^3(\chi
   _0) \sin (2 (\phi +2 \chi _0+\tan ^{-1}(\sqrt{2}))) t_f t_{2 \phi } t_{\text{ps}}^5
   \\
   -64 \cos (\chi
   _0) \sin ^3(\chi _0) \sin (2 (\phi +3 \chi _0+\tan ^{-1}(\sqrt{2}))) t_f t_{2 \phi }
   t_{\text{ps}}^5
   \\
   -128 \cos (\chi _0) \sin ^3(\chi _0) \sin (2 \phi +4 \chi _0-2 \tan ^{-1}(\sqrt{2})) t_f
   t_{2 \phi } t_{\text{ps}}^5
   \\
   -64 \cos (\chi _0) \sin ^3(\chi _0) \sin (2 \phi +6 \chi _0-2 \tan
   ^{-1}(\sqrt{2})) t_f t_{2 \phi } t_{\text{ps}}^5+112 \cos (\phi ) \sin (\chi _0) t_f t_{\phi } t_{2 \phi }
   t_{\text{ps}}^5
   \\
   -4 \cos (\phi -10 \chi _0) \sin (\chi _0) t_f t_{\phi } t_{2 \phi } t_{\text{ps}}^5-24 \cos (\phi -8 \chi
   _0) \sin (\chi _0) t_f t_{\phi } t_{2 \phi } t_{\text{ps}}^5-52 \cos (\phi -6 \chi _0) \sin (\chi _0) t_f
   t_{\phi } t_{2 \phi } t_{\text{ps}}^5
   \\
   -32 \cos (\phi -4 \chi _0) \sin (\chi _0) t_f t_{\phi } t_{2 \phi } t_{\text{ps}}^5+56
   \cos (\phi -2 \chi _0) \sin (\chi _0) t_f t_{\phi } t_{2 \phi } t_{\text{ps}}^5+56 \cos (\phi +2 \chi _0) \sin
   (\chi _0) t_f t_{\phi } t_{2 \phi } t_{\text{ps}}^5
   \\
   -32 \cos (\phi +4 \chi _0) \sin (\chi _0) t_f t_{\phi } t_{2
   \phi } t_{\text{ps}}^5-52 \cos (\phi +6 \chi _0) \sin (\chi _0) t_f t_{\phi } t_{2 \phi } t_{\text{ps}}^5-24 \cos (\phi
   +8 \chi _0) \sin (\chi _0) t_f t_{\phi } t_{2 \phi } t_{\text{ps}}^5
   \\
   -4 \cos (\phi +10 \chi _0) \sin (\chi _0)
   t_f t_{\phi } t_{2 \phi } t_{\text{ps}}^5-12 \cos (6 \chi _0) t_{\phi }^2 t_{\text{ps}}^4+2 \cos (8 \chi _0) t_{\phi }^2
   t_{\text{ps}}^4+4 \cos (10 \chi _0) t_{\phi }^2 t_{\text{ps}}^4+\cos (12 \chi _0) t_{\phi }^2 t_{\text{ps}}^4
   \\
   -1280 \sin
   (\text{$\Delta $x}_4+\tan ^{-1}(\sqrt{2})) \sin ^2(\chi _0) t_f^2 t_{2 \phi }^2 t_{\text{ps}}^4-64 \sin
   (\text{$\Delta $x}_4-6 \chi _0+\tan ^{-1}(\sqrt{2})) \sin ^2(\chi _0) t_f^2 t_{2 \phi }^2 t_{\text{ps}}^4
   \\
   -384 \sin
   (\text{$\Delta $x}_4-4 \chi _0+\tan ^{-1}(\sqrt{2})) \sin ^2(\chi _0) t_f^2 t_{2 \phi }^2 t_{\text{ps}}^4-960 \sin
   (\text{$\Delta $x}_4-2 \chi _0+\tan ^{-1}(\sqrt{2})) \sin ^2(\chi _0) t_f^2 t_{2 \phi }^2 t_{\text{ps}}^4
   \\
   -960 \sin
   ^2(\chi _0) \sin (\text{$\Delta $x}_4+2 \chi _0+\tan ^{-1}(\sqrt{2})) t_f^2 t_{2 \phi }^2 t_{\text{ps}}^4-384 \sin
   ^2(\chi _0) \sin (\text{$\Delta $x}_4+4 \chi _0+\tan ^{-1}(\sqrt{2})) t_f^2 t_{2 \phi }^2 t_{\text{ps}}^4
   \\
   -64 \sin
   ^2(\chi _0) \sin (\text{$\Delta $x}_4+6 \chi _0+\tan ^{-1}(\sqrt{2})) t_f^2 t_{2 \phi }^2 t_{\text{ps}}^4-12 \cos
   (6 \chi _0) t_{\text{ps}}^4+8 \cos (8 \chi _0) t_{\text{ps}}^4+4 \cos (10 \chi _0) t_{\text{ps}}^4
   \\
   +96 \cos
   (\chi _0) \sin (\phi -2 \tan ^{-1}(\sqrt{2})) \sin ^2(\chi _0) t_{\phi } t_{\text{ps}}^4+64 \cos
   (\chi _0) \sin (\phi -6 \chi _0+2 \tan ^{-1}(\sqrt{2})) \sin ^2(\chi _0) t_{\phi } t_{\text{ps}}^4
   \\
   -64
   \cos (\chi _0) \sin (\phi -4 \chi _0-2 \tan ^{-1}(\sqrt{2})) \sin ^2(\chi _0) t_{\phi }
   t_{\text{ps}}^4+64 \cos (\chi _0) \sin (\phi -4 \chi _0+2 \tan ^{-1}(\sqrt{2})) \sin ^2(\chi _0) t_{\phi
   } t_{\text{ps}}^4
   \\
   -64 \cos (\chi _0) \sin (\phi -2 \chi _0+2 \tan ^{-1}(\sqrt{2})) \sin ^2(\chi _0)
   t_{\phi } t_{\text{ps}}^4-80 \sin (\phi +2 \tan ^{-1}(\sqrt{2})) \sin (\chi _0) \sin (2 \chi _0) t_{\phi
   } t_{\text{ps}}^4
   \\
   +8 \sin (\phi -8 \chi _0-2 \tan ^{-1}(\sqrt{2})) \sin (\chi _0) \sin (2 \chi _0)
   t_{\phi } t_{\text{ps}}^4+8 \sin (\phi -8 \chi _0+2 \tan ^{-1}(\sqrt{2})) \sin (\chi _0) \sin (2 \chi
   _0) t_{\phi } t_{\text{ps}}^4
   \\
   -64 \cos (\chi _0) \sin ^2(\chi _0) \sin (\phi +4 \chi _0-2 \tan
   ^{-1}(\sqrt{2})) t_{\phi } t_{\text{ps}}^4+64 \cos (\chi _0) \sin ^2(\chi _0) \sin (\phi +4 \chi _0+2
   \tan ^{-1}(\sqrt{2})) t_{\phi } t_{\text{ps}}^4
   \\
   +64 \cos (\chi _0) \sin ^2(\chi _0) \sin (\phi +6 \chi
   _0+2 \tan ^{-1}(\sqrt{2})) t_{\phi } t_{\text{ps}}^4+8 \sin (\chi _0) \sin (2 \chi _0) \sin (\phi +8
   \chi _0-2 \tan ^{-1}(\sqrt{2})) t_{\phi } t_{\text{ps}}^4
   \\
   +8 \sin (\chi _0) \sin (2 \chi _0) \sin (\phi
   +8 \chi _0+2 \tan ^{-1}(\sqrt{2})) t_{\phi } t_{\text{ps}}^4-64 \cos (\chi _0) \sin ^2(\chi _0) \sin
   (\phi +2 (\chi _0+\tan ^{-1}(\sqrt{2}))) t_{\phi } t_{\text{ps}}^4
   \\
   +2560 \cos (\phi ) \cos (\chi _0) \sin
   (\phi ) \sin ^3(\chi _0) t_f t_{2 \phi } t_{\text{ps}}^4+64 \cos (\chi _0) \sin (2 \phi -6 \chi _0) \sin
   ^3(\chi _0) t_f t_{2 \phi } t_{\text{ps}}^4
   \\
   +384 \cos (\chi _0) \sin (2 \phi -4 \chi _0) \sin ^3(\chi _0)
   t_f t_{2 \phi } t_{\text{ps}}^4+960 \cos (\chi _0) \sin (2 \phi -2 \chi _0) \sin ^3(\chi _0) t_f t_{2 \phi }
   t_{\text{ps}}^4
   \\
   +960 \cos (\chi _0) \sin ^3(\chi _0) \sin (2 (\phi +\chi _0)) t_f t_{2 \phi }
   t_{\text{ps}}^4+384 \cos (\chi _0) \sin ^3(\chi _0) \sin (2 \phi +4 \chi _0) t_f t_{2 \phi } t_{\text{ps}}^4
   \\
   +64
   \cos (\chi _0) \sin ^3(\chi _0) \sin (2 \phi +6 \chi _0) t_f t_{2 \phi } t_{\text{ps}}^4+96 \sin (\phi
   -\text{$\Delta $x}_4+\tan ^{-1}(\sqrt{2})) \sin (\chi _0) t_f t_{\phi } t_{2 \phi } t_{\text{ps}}^4
   \\
   +160 \sin (\phi
   +\text{$\Delta $x}_4-\tan ^{-1}(\sqrt{2})) \sin (\chi _0) t_f t_{\phi } t_{2 \phi } t_{\text{ps}}^4+16 \sin (\phi
   -\text{$\Delta $x}_4-8 \chi _0+\tan ^{-1}(\sqrt{2})) \sin (\chi _0) t_f t_{\phi } t_{2 \phi } t_{\text{ps}}^4
   \\
   -16 \sin
   (\phi +\text{$\Delta $x}_4-8 \chi _0-\tan ^{-1}(\sqrt{2})) \sin (\chi _0) t_f t_{\phi } t_{2 \phi }
   t_{\text{ps}}^4+64 \sin (\phi +\text{$\Delta $x}_4-6 \chi _0-\tan ^{-1}(\sqrt{2})) \sin (\chi _0) t_f t_{\phi }
   t_{2 \phi } t_{\text{ps}}^4
   \\
   -64 \sin (\phi -\text{$\Delta $x}_4-4 \chi _0+\tan ^{-1}(\sqrt{2})) \sin (\chi _0) t_f
   t_{\phi } t_{2 \phi } t_{\text{ps}}^4-64 \sin (\phi +\text{$\Delta $x}_4-4 \chi _0-\tan ^{-1}(\sqrt{2})) \sin (\chi
   _0) t_f t_{\phi } t_{2 \phi } t_{\text{ps}}^4
   \\
   -64 \sin (\phi +\text{$\Delta $x}_4-2 \chi _0-\tan ^{-1}(\sqrt{2})) \sin
   (\chi _0) t_f t_{\phi } t_{2 \phi } t_{\text{ps}}^4-64 \sin (\chi _0) \sin (\phi +\text{$\Delta $x}_4+2 \chi _0-\tan
   ^{-1}(\sqrt{2})) t_f t_{\phi } t_{2 \phi } t_{\text{ps}}^4
   \\
   -64 \sin (\chi _0) \sin (\phi -\text{$\Delta $x}_4+4 \chi
   _0+\tan ^{-1}(\sqrt{2})) t_f t_{\phi } t_{2 \phi } t_{\text{ps}}^4-64 \sin (\chi _0) \sin (\phi +\text{$\Delta
   $x}_4+4 \chi _0-\tan ^{-1}(\sqrt{2})) t_f t_{\phi } t_{2 \phi } t_{\text{ps}}^4
   \\
   +64 \sin (\chi _0) \sin (\phi
   +\text{$\Delta $x}_4+6 \chi _0-\tan ^{-1}(\sqrt{2})) t_f t_{\phi } t_{2 \phi } t_{\text{ps}}^4+16 \sin (\chi _0) \sin
   (\phi -\text{$\Delta $x}_4+8 \chi _0+\tan ^{-1}(\sqrt{2})) t_f t_{\phi } t_{2 \phi } t_{\text{ps}}^4
   \\
   -16 \sin (\chi
   _0) \sin (\phi +\text{$\Delta $x}_4+8 \chi _0-\tan ^{-1}(\sqrt{2})) t_f t_{\phi } t_{2 \phi } t_{\text{ps}}^4-112 \sin
   (\tan ^{-1}(\sqrt{2})-\text{$\Delta $x}_4) t_{\phi }^2 t_{\text{ps}}^3
   \\
   -4 \sin (-\text{$\Delta $x}_4-10 \chi _0+\tan
   ^{-1}(\sqrt{2})) t_{\phi }^2 t_{\text{ps}}^3+24 \sin (-\text{$\Delta $x}_4-8 \chi _0+\tan ^{-1}(\sqrt{2}))
   t_{\phi }^2 t_{\text{ps}}^3
   \\
   -52 \sin (-\text{$\Delta $x}_4-6 \chi _0+\tan ^{-1}(\sqrt{2})) t_{\phi }^2 t_{\text{ps}}^3+32 \sin
   (-\text{$\Delta $x}_4-4 \chi _0+\tan ^{-1}(\sqrt{2})) t_{\phi }^2 t_{\text{ps}}^3
   \\
   +56 \sin (-\text{$\Delta $x}_4-2 \chi
   _0+\tan ^{-1}(\sqrt{2})) t_{\phi }^2 t_{\text{ps}}^3+56 \sin (-\text{$\Delta $x}_4+2 \chi _0+\tan
   ^{-1}(\sqrt{2})) t_{\phi }^2 t_{\text{ps}}^3
   \\
   +32 \sin (-\text{$\Delta $x}_4+4 \chi _0+\tan ^{-1}(\sqrt{2}))
   t_{\phi }^2 t_{\text{ps}}^3-52 \sin (-\text{$\Delta $x}_4+6 \chi _0+\tan ^{-1}(\sqrt{2})) t_{\phi }^2 t_{\text{ps}}^3
   \\
   +24 \sin
   (-\text{$\Delta $x}_4+8 \chi _0+\tan ^{-1}(\sqrt{2})) t_{\phi }^2 t_{\text{ps}}^3-4 \sin (-\text{$\Delta $x}_4+10 \chi
   _0+\tan ^{-1}(\sqrt{2})) t_{\phi }^2 t_{\text{ps}}^3-4 \cos (2 \tan ^{-1}(\sqrt{2})-10 \chi _0)
   t_{\text{ps}}^3
   \\
   -8 \cos (2 \tan ^{-1}(\sqrt{2})-8 \chi _0) t_{\text{ps}}^3+12 \cos (2 \tan ^{-1}(\sqrt{2})-6
   \chi _0) t_{\text{ps}}^3+32 \cos (2 \tan ^{-1}(\sqrt{2})-4 \chi _0) t_{\text{ps}}^3
   \\
   -8 \cos (2 \tan
   ^{-1}(\sqrt{2})-2 \chi _0) t_{\text{ps}}^3-8 \cos (2 (\chi _0+\tan ^{-1}(\sqrt{2})))
   t_{\text{ps}}^3+32 \cos (4 \chi _0+2 \tan ^{-1}(\sqrt{2})) t_{\text{ps}}^3
   \\
   -4 \cos (2 (5 \chi _0+\tan
   ^{-1}(\sqrt{2}))) t_{\text{ps}}^3+12 \cos (6 \chi _0+2 \tan ^{-1}(\sqrt{2})) t_{\text{ps}}^3-8 \cos
   (8 \chi _0+2 \tan ^{-1}(\sqrt{2})) t_{\text{ps}}^3
   \\
   -64 \cos (\chi _0) \sin (\phi -6 \chi _0) \sin
   ^2(\chi _0) t_{\phi } t_{\text{ps}}^3-64 \cos (\chi _0) \sin (\phi -4 \chi _0) \sin ^2(\chi _0) t_{\phi
   } t_{\text{ps}}^3+64 \cos (\chi _0) \sin (\phi -2 \chi _0) \sin ^2(\chi _0) t_{\phi } t_{\text{ps}}^3
   \\
   +80 \sin (\phi
   ) \sin (\chi _0) \sin (2 \chi _0) t_{\phi } t_{\text{ps}}^3-8 \sin (\phi -8 \chi _0) \sin (\chi _0) \sin
   (2 \chi _0) t_{\phi } t_{\text{ps}}^3+64 \cos (\chi _0) \sin ^2(\chi _0) \sin (\phi +2 \chi _0) t_{\phi
   } t_{\text{ps}}^3
   \\
   -64 \cos (\chi _0) \sin ^2(\chi _0) \sin (\phi +4 \chi _0) t_{\phi } t_{\text{ps}}^3-64 \cos
   (\chi _0) \sin ^2(\chi _0) \sin (\phi +6 \chi _0) t_{\phi } t_{\text{ps}}^3-8 \sin (\chi _0) \sin
   (2 \chi _0) \sin (\phi +8 \chi _0) t_{\phi } t_{\text{ps}}^3
   \\
   +1536 \cos (2 \phi -\text{$\Delta $x}_4+\tan
   ^{-1}(\sqrt{2})) \cos (\chi _0) \sin ^3(\chi _0) t_f t_{2 \phi } t_{\text{ps}}^3
   \\
   +256 \cos (2 \phi
   -\text{$\Delta $x}_4-4 \chi _0+\tan ^{-1}(\sqrt{2})) \cos (\chi _0) \sin ^3(\chi _0) t_f t_{2 \phi }
   t_{\text{ps}}^3
   \\
   +1024 \cos (2 \phi -\text{$\Delta $x}_4-2 \chi _0+\tan ^{-1}(\sqrt{2})) \cos (\chi _0) \sin
   ^3(\chi _0) t_f t_{2 \phi } t_{\text{ps}}^3
   \\
   +256 \cos (\chi _0) \cos (2 \phi -\text{$\Delta $x}_4+4 \chi _0+\tan
   ^{-1}(\sqrt{2})) \sin ^3(\chi _0) t_f t_{2 \phi } t_{\text{ps}}^3
   \\
   +1024 \cos (\chi _0) \cos
   (-\text{$\Delta $x}_4+2 (\phi +\chi _0)+\tan ^{-1}(\sqrt{2})) \sin ^3(\chi _0) t_f t_{2 \phi }
   t_{\text{ps}}^3
   \\
   +160 \sin (\phi -\text{$\Delta $x}_4-\tan ^{-1}(\sqrt{2})) \sin (\chi _0) t_f t_{\phi } t_{2 \phi }
   t_{\text{ps}}^3+96 \sin (\phi +\text{$\Delta $x}_4+\tan ^{-1}(\sqrt{2})) \sin (\chi _0) t_f t_{\phi } t_{2 \phi }
   t_{\text{ps}}^3
   \\
   -16 \sin (\phi -\text{$\Delta $x}_4-8 \chi _0-\tan ^{-1}(\sqrt{2})) \sin (\chi _0) t_f t_{\phi }
   t_{2 \phi } t_{\text{ps}}^3+16 \sin (\phi +\text{$\Delta $x}_4-8 \chi _0+\tan ^{-1}(\sqrt{2})) \sin (\chi _0) t_f
   t_{\phi } t_{2 \phi } t_{\text{ps}}^3
   \\
   -64 \sin (\phi -\text{$\Delta $x}_4-6 \chi _0-\tan ^{-1}(\sqrt{2})) \sin (\chi
   _0) t_f t_{\phi } t_{2 \phi } t_{\text{ps}}^3-64 \sin (\phi -\text{$\Delta $x}_4-4 \chi _0-\tan ^{-1}(\sqrt{2})) \sin
   (\chi _0) t_f t_{\phi } t_{2 \phi } t_{\text{ps}}^3
   \\
   -64 \sin (\phi +\text{$\Delta $x}_4-4 \chi _0+\tan
   ^{-1}(\sqrt{2})) \sin (\chi _0) t_f t_{\phi } t_{2 \phi } t_{\text{ps}}^3+64 \sin (\phi -\text{$\Delta $x}_4-2 \chi
   _0-\tan ^{-1}(\sqrt{2})) \sin (\chi _0) t_f t_{\phi } t_{2 \phi } t_{\text{ps}}^3
   \\
   +64 \sin (\chi _0) \sin
   (\phi -\text{$\Delta $x}_4+2 \chi _0-\tan ^{-1}(\sqrt{2})) t_f t_{\phi } t_{2 \phi } t_{\text{ps}}^3-64 \sin (\chi
   _0) \sin (\phi -\text{$\Delta $x}_4+4 \chi _0-\tan ^{-1}(\sqrt{2})) t_f t_{\phi } t_{2 \phi } t_{\text{ps}}^3
   \\
   -64 \sin
   (\chi _0) \sin (\phi +\text{$\Delta $x}_4+4 \chi _0+\tan ^{-1}(\sqrt{2})) t_f t_{\phi } t_{2 \phi }
   t_{\text{ps}}^3-64 \sin (\chi _0) \sin (\phi -\text{$\Delta $x}_4+6 \chi _0-\tan ^{-1}(\sqrt{2})) t_f t_{\phi }
   t_{2 \phi } t_{\text{ps}}^3
   \\
   -16 \sin (\chi _0) \sin (\phi -\text{$\Delta $x}_4+8 \chi _0-\tan ^{-1}(\sqrt{2})) t_f
   t_{\phi } t_{2 \phi } t_{\text{ps}}^3+16 \sin (\chi _0) \sin (\phi +\text{$\Delta $x}_4+8 \chi _0+\tan
   ^{-1}(\sqrt{2})) t_f t_{\phi } t_{2 \phi } t_{\text{ps}}^3
   \\
   +48 \sin (\text{$\Delta $x}_4+\tan ^{-1}(\sqrt{2}))
   t_{\phi }^2 t_{\text{ps}}^2-4 \sin (\text{$\Delta $x}_4-10 \chi _0+\tan ^{-1}(\sqrt{2})) t_{\phi }^2 t_{\text{ps}}^2+8 \sin
   (\text{$\Delta $x}_4-8 \chi _0+\tan ^{-1}(\sqrt{2})) t_{\phi }^2 t_{\text{ps}}^2
   \\
   +12 \sin (\text{$\Delta $x}_4-6 \chi
   _0+\tan ^{-1}(\sqrt{2})) t_{\phi }^2 t_{\text{ps}}^2-32 \sin (\text{$\Delta $x}_4-4 \chi _0+\tan
   ^{-1}(\sqrt{2})) t_{\phi }^2 t_{\text{ps}}^2
   \\
   -8 \sin (\text{$\Delta $x}_4-2 \chi _0+\tan ^{-1}(\sqrt{2}))
   t_{\phi }^2 t_{\text{ps}}^2-8 \sin (\text{$\Delta $x}_4+2 \chi _0+\tan ^{-1}(\sqrt{2})) t_{\phi }^2 t_{\text{ps}}^2
   \\
   -32 \sin
   (\text{$\Delta $x}_4+4 \chi _0+\tan ^{-1}(\sqrt{2})) t_{\phi }^2 t_{\text{ps}}^2+12 \sin (\text{$\Delta $x}_4+6 \chi
   _0+\tan ^{-1}(\sqrt{2})) t_{\phi }^2 t_{\text{ps}}^2
   \\
   +8 \sin (\text{$\Delta $x}_4+8 \chi _0+\tan
   ^{-1}(\sqrt{2})) t_{\phi }^2 t_{\text{ps}}^2-4 \sin (\text{$\Delta $x}_4+10 \chi _0+\tan ^{-1}(\sqrt{2}))
   t_{\phi }^2 t_{\text{ps}}^2+768 \sin ^2(\chi _0) t_f^2 t_{2 \phi }^2 t_{\text{ps}}^2
   \\
   -12 \cos (6 \chi _0) t_{\text{ps}}^2+8
   \cos (8 \chi _0) t_{\text{ps}}^2+4 \cos (10 \chi _0) t_{\text{ps}}^2+8 (t_{\text{ps}} (3
   t_{\text{ps}}+2)+3) t_{\text{ps}}^2
   \\
   -256 \cos (\phi -\text{$\Delta $x}_4+\tan ^{-1}(\sqrt{2})) \cos (\chi
   _0) \sin ^2(\chi _0) t_{\phi } t_{\text{ps}}^2-64 \cos (\phi -\text{$\Delta $x}_4-6 \chi _0+\tan
   ^{-1}(\sqrt{2})) \cos (\chi _0) \sin ^2(\chi _0) t_{\phi } t_{\text{ps}}^2
   \\
   +128 \cos (\phi -\text{$\Delta
   $x}_4-4 \chi _0+\tan ^{-1}(\sqrt{2})) \cos (\chi _0) \sin ^2(\chi _0) t_{\phi } t_{\text{ps}}^2
   \\
   +64 \cos
   (\phi -\text{$\Delta $x}_4-2 \chi _0+\tan ^{-1}(\sqrt{2})) \cos (\chi _0) \sin ^2(\chi _0) t_{\phi }
   t_{\text{ps}}^2
   \\
   +64 \cos (\chi _0) \cos (\phi -\text{$\Delta $x}_4+2 \chi _0+\tan ^{-1}(\sqrt{2})) \sin ^2(\chi
   _0) t_{\phi } t_{\text{ps}}^2
   \\
   +128 \cos (\chi _0) \cos (\phi -\text{$\Delta $x}_4+4 \chi _0+\tan
   ^{-1}(\sqrt{2})) \sin ^2(\chi _0) t_{\phi } t_{\text{ps}}^2
   \\
   -64 \cos (\chi _0) \cos (\phi -\text{$\Delta
   $x}_4+6 \chi _0+\tan ^{-1}(\sqrt{2})) \sin ^2(\chi _0) t_{\phi } t_{\text{ps}}^2
   \\
   -1536 \cos (2 \phi -\text{$\Delta
   $x}_4-\tan ^{-1}(\sqrt{2})) \cos (\chi _0) \sin ^3(\chi _0) t_f t_{2 \phi } t_{\text{ps}}^2
   \\
   -256 \cos (2
   \phi -\text{$\Delta $x}_4-4 \chi _0-\tan ^{-1}(\sqrt{2})) \cos (\chi _0) \sin ^3(\chi _0) t_f t_{2 \phi }
   t_{\text{ps}}^2
   \\
   -1024 \cos (2 \phi -\text{$\Delta $x}_4-2 \chi _0-\tan ^{-1}(\sqrt{2})) \cos (\chi _0) \sin
   ^3(\chi _0) t_f t_{2 \phi } t_{\text{ps}}^2
   \\
   -256 \cos (\chi _0) \cos (2 \phi -\text{$\Delta $x}_4+4 \chi _0-\tan
   ^{-1}(\sqrt{2})) \sin ^3(\chi _0) t_f t_{2 \phi } t_{\text{ps}}^2
   \\
   -1024 \cos (\chi _0) \cos
   (\text{$\Delta $x}_4-2 (\phi +\chi _0)+\tan ^{-1}(\sqrt{2})) \sin ^3(\chi _0) t_f t_{2 \phi }
   t_{\text{ps}}^2
   \\
   +256 \cos (\phi -\text{$\Delta $x}_4-\tan ^{-1}(\sqrt{2})) \cos (\chi _0) \sin ^2(\chi
   _0) t_{\phi } t_{\text{ps}}
   \\
   +64 \cos (\phi -\text{$\Delta $x}_4-6 \chi _0-\tan ^{-1}(\sqrt{2})) \cos (\chi _0)
   \sin ^2(\chi _0) t_{\phi } t_{\text{ps}}
   \\
   -128 \cos (\phi -\text{$\Delta $x}_4-4 \chi _0-\tan ^{-1}(\sqrt{2})) \cos
   (\chi _0) \sin ^2(\chi _0) t_{\phi } t_{\text{ps}}
   \\
   -64 \cos (\phi -\text{$\Delta $x}_4-2 \chi _0-\tan
   ^{-1}(\sqrt{2})) \cos (\chi _0) \sin ^2(\chi _0) t_{\phi } t_{\text{ps}}
   \\
   -64 \cos (\chi _0) \cos
   (\phi -\text{$\Delta $x}_4+2 \chi _0-\tan ^{-1}(\sqrt{2})) \sin ^2(\chi _0) t_{\phi } t_{\text{ps}}
   \\
   -128 \cos
   (\chi _0) \cos (\phi -\text{$\Delta $x}_4+4 \chi _0-\tan ^{-1}(\sqrt{2})) \sin ^2(\chi _0) t_{\phi }
   t_{\text{ps}}
   \\
   +64 \cos (\chi _0) \cos (\phi -\text{$\Delta $x}_4+6 \chi _0-\tan ^{-1}(\sqrt{2})) \sin ^2(\chi
   _0) t_{\phi } t_{\text{ps}}
   \\
   -256 \cos (\phi ) \sin (\chi _0) t_f t_{\phi } t_{2 \phi } t_{\text{ps}}-64 \cos (\phi -6 \chi
   _0) \sin (\chi _0) t_f t_{\phi } t_{2 \phi } t_{\text{ps}}
   \\
   +128 \cos (\phi -4 \chi _0) \sin (\chi _0) t_f
   t_{\phi } t_{2 \phi } t_{\text{ps}}+64 \cos (\phi -2 \chi _0) \sin (\chi _0) t_f t_{\phi } t_{2 \phi } t_{\text{ps}}
   \\
   +64 \cos
   (\phi +2 \chi _0) \sin (\chi _0) t_f t_{\phi } t_{2 \phi } t_{\text{ps}}+128 \cos (\phi +4 \chi _0) \sin (\chi
   _0) t_f t_{\phi } t_{2 \phi } t_{\text{ps}}-64 \cos (\phi +6 \chi _0) \sin (\chi _0) t_f t_{\phi } t_{2 \phi }
   t_{\text{ps}}
   \\
   -128 \cos (6 \chi _0) t_{\phi }^2+16 \cos (8 \chi _0) t_{\phi }^2+\frac{2}{3} ((t_{\text{ps}} (21
   t_{\text{ps}}-10)+21) t_{\text{ps}}^4+840) t_{\phi }^2
   \\
   +8 \cos (2 \chi _0) (t_{\text{ps}}^4+16 \sin ^2(\chi
   _0) t_f^2 (7 t_{\text{ps}}^4+8) t_{2 \phi }^2 t_{\text{ps}}^2+t_{\text{ps}}^2+(-t_{\text{ps}}^6+t_{\text{ps}}^4-112)
   t_{\phi }^2)+\cos (4 \chi _0) ((448-17 (t_{\text{ps}}^6+t_{\text{ps}}^4)) t_{\phi }^2
   \\
   -64 \sin
   ^2(\chi _0) t_f^2 t_{\text{ps}}^2 (t_{\text{ps}}^6-7 t_{\text{ps}}^4-4) t_{2 \phi }^2-32
   (t_{\text{ps}}^4+t_{\text{ps}}^2)))
\end{multline}
\endgroup

For $\Delta x_4 = 0$ this becomes $\tilde{p}_4 = \min_{x_4} \tilde{p}_4 + 0.64\,(\max_{x_4} \tilde{p}_4-\min_{x_4} \tilde{p}_4)$ (see Fig.\,\ref{adj}(d)).



\section*{Scaling limitations}
In this section we discuss the main limitations for the scalability of optical multiport schemes with the architecture similar to that presented in the paper.
Below is the list of the main scalability-detrimental factors.\\
    \textit{Restricted phase adjustment precision.--} 
    An $N$-dimensional unitary matrix $U$ can be decomposed into the product of $M=N(N-1)/2$ two-level matrices: 
    \begin{equation}
    U = \prod_{i=1}^M T_i,
    \end{equation}
    with 
    \begin{equation}
    T_i = \begin{pmatrix}
    \cos{(\phi_i/2)} & i \sin{(\phi_i/2)} \\
    i\sin{(\phi_i/2)} & \cos{(\phi_i/2)}
    \end{pmatrix},
    \end{equation}
    where $\phi_i$ is the phase determining the split-ratio of the beam splitter.
    Our setup includes only symmetric beam splitters. 
    However, one can effectively realize an arbitrary split-ratio by utilizing Mach--Zehnder interferometry.
    This yields the phase $\phi_i$ ($i\in\{1,\,\dots,\,M\}$) adjustment with the precision $\Delta \phi$. 
    We can thus write the total cumulative error in terms of the matrix norms:
    \begin{equation}
        \lVert \Delta U \rVert = \lVert \sum_{i=1}^M T_1...T_{i-1} \Delta T_i T_{i+1} ... T_M \rVert \leq \prod_i^M \lVert T_i \rVert \cdot \left( \sum_{i=1}^M \lVert \frac{\Delta T_i}{T_i}\rVert \right).
    \end{equation}
    The logarithmic derivative
    \begin{equation}
    \frac{\Delta T_i/T_i}{\Delta \phi_i} = \frac{1}{2} \begin{pmatrix}
    0 & i \\
    i & 0
    \end{pmatrix}.
    \end{equation}
    Conversely, a crude estimation gives
    \begin{equation}
    \frac{\lVert \Delta U \rVert}{\prod_{i=1}^M \lVert T_i \rVert}  \leq \sum_{i=1}^M \lVert \frac{\Delta T_i}{T_i} \rVert \sim  \sqrt{M} \Delta \phi \sim N \Delta \phi.
    \end{equation}
    In the experiment we had $\Delta \phi \approx 0.2 \div 0.3$.
    That said, we expect that the use of thinner glass phase shifters with width $d=0.1$\,mm and optical holders with finer control precision $\Delta \alpha \sim 10^{-5}$ ($\alpha$ is the rotation angle of a holder) should enable $\Delta \phi = 10^{-3}$:
    \begin{equation}
        \Delta \phi = \left( \frac{\partial \phi}{\partial \alpha} \right) \Delta \alpha \approx
        \frac{2\pi (n-1) d}{\lambda} \frac{\sin{\alpha}}{\cos{\alpha}^2} \Delta \alpha \approx 10^{-3},
    \end{equation}
    where $n$ is the refractive index of the phase shifter, $\lambda$ is the light wavelength.
    \\
    \textit{Restricted precision of the wavefronts' alignment.--}
    Let us estimate the error caused by the misalignment of the wavefronts.
    The signal intensities for the circular- ($I_\bigcirc$) and square-shaped ($I_\square$) beams are given by
    \begin{equation}
        I_\bigcirc =
        \iint_S \frac{d^2 \mathbf{r}}{\pi R^2} \lvert \sum_{j=1}^K \frac{1}{\sqrt{K}} e^{i (\mathbf{k}_j \mathbf{r} - \xi_j)}\rvert^2  = 
        \frac{1}{K} \sum_{i,j=1}^K e^{i(\xi_j - \xi_i)} \frac{2 J_1 ( \lvert \mathbf{k}_i - \mathbf{k}_j \rvert R )}{\lvert \mathbf{k}_i - \mathbf{k}_j \rvert R} \sim 
    1 + I^0_\bigcirc(\xi_1,\,\xi_2,\,\dots) \frac{2 J_1 ( \Delta k R )}{\Delta k R},
    \end{equation}
    \begin{multline}
        I_\square =
        \iint_S \frac{d^2 \mathbf{r}}{a^2} \lvert \sum_{j=1}^K \frac{1}{\sqrt{K}} e^{i (\mathbf{k}_j \mathbf{r} - \xi_j)}\rvert^2  = 
        \frac{1}{K} \sum_{i,j=1}^K e^{i(\xi_j - \xi_i)} \frac{\sin{\left( (k_i^{(x)}-k_j^{(x)}) a \right) }}{(k_i^{(x)}-k_j^{(x)}) a} \frac{\sin{\left( (k_i^{(y)}-k_j^{(y)}) a \right) }}{(k_i^{(y)}-k_j^{(y)}) a} \\\sim 
        1 + I^0_\square(\xi_1,\,\xi_2,\,\dots) \frac{\sin{\Delta k a}}{\Delta k a},
    \end{multline}
    where the summation is performed over all different trajectories along which the light can travel to the detector through the scheme, the subscript $i$ indicates the number of the trajectory,
    $K$ is the total number of trajectories incident on the detector which grows exponentially with $N$,
    $\mathbf{k}_i$ is the wave vector component parallel to the detector's surface, $\xi_i$ is the phase in the center, $\Delta k$ is the characteristic variation of the wave vectors, $R$ and $a$ are, respectively, the radius of the circular-shaped beam and the side length of the square-shaped beam, $J_1$ is the first order Bessel function, $I^0_\bigcirc(\xi_1,\,\xi_2,\,\dots)$ and $I^0_\square(\xi_1,\,\xi_2,\,\dots)$ are the phase-depended factors.
    In the ideal case where the wavefronts' misalignment is absent, we have
    \begin{equation}
    I_{\square(\bigcirc)} \sim 
    1 + I^0_{\square(\bigcirc)} (\xi_1,\,\xi_2,\,\dots).
    \end{equation}
    Thus, the deterioration of the interference picture due to the wavefronts' misalignment is reflected in the factors $\frac{2 J_1 ( \Delta k R )}{\Delta k R}$ and $\frac{\sin{\Delta k a}}{\Delta k a}$.
    
    Although a scheme realizing an $N$-dimensional unitary matrix comprises of order $N^2$ beam splitters, each trajectory passing only through $\sim N$ of them.
    Supposing that on passing the $j$th beam splitter the wave vector $\mathbf{k}_i$ diverges for $\mathbf{q}_{i,j}$, and the typical length of $\mathbf{q}_{i,j}$ is $\Delta q$, we get 
    \begin{equation}
        \Delta k \sim \sqrt{N} \Delta q.
    \end{equation}
    Here we also assumed that for any $j$ and $m$ such that $j\neq m$, $\mathbf{q}_{i,j}$ and $\mathbf{q}_{i,m}$ are independent.
    The visibility spoils linearly with the increase of $N$:
    \begin{gather}
    \frac{2 J_1 ( \Delta k R )}{\Delta k R} \approx
    1 - \frac{N(\Delta q R)^2}{8};\\
    \frac{\sin{\Delta k a}}{\Delta k a} \approx
    1 - \frac{N(\Delta q a)^2}{6}. 
    \end{gather}
    If the angle of optical holders is adjusted with the precision $\Delta \alpha$, then $\Delta q = \Delta \alpha \frac{2\pi}{\lambda}$.
    The equipment employed in our experiment allows for $\Delta \alpha \sim 10^{-5}$ and $R\simeq1$\,mm (or $a=2$\,mm); thus, $\frac{2 J_1 ( \Delta k R )}{\Delta k R}\approx
    1 - \frac{N\pi^2}{5000}$ and $\frac{\sin{\Delta k a}}{\Delta k a}  \approx
    1 - \frac{N\pi^2}{1000}$. One sees that decreasing $R$ or $a$ (which can be easily achieved by the decreasing the detector area), one can further improve the precision.
    \\
    \textit{Phase fluctuations caused by the surface roughness.--} 
    Assuming that the light acquires delta-correlated random phase $\delta \xi_j(\mathbf{r})$ due to the surface roughness of the optical elements, we can, in a manner similar to the above, write a relation for the signal intensity:
    \begin{equation}
    I =
    \langle \iint_S \frac{d^2 \mathbf{r}}{\pi R^2} \lvert \sum_{j=1}^K \frac{1}{\sqrt{K}} e^{i (\delta \xi_j(\mathbf{r}) - \xi_j)}\rvert^2  \rangle = 
    \frac{1}{K} \sum_{i,j=1}^K e^{i(\xi_j - \xi_i)} 
    \iint_S \frac{d^2 \mathbf{r}}{\pi R^2} \langle e^{i(\delta \xi_i(\mathbf{r}) - \delta \xi_j(\mathbf{r}))} \rangle
    \sim 
    1 + I_0 e^{-N \langle \delta \xi^2 \rangle},
    \end{equation}
    where averaging $\langle\dots\rangle$ is done over different random phases; we assumed that the beams have a circular shape.
    We estimate that in our setup $\delta \xi \sim 2\pi / 100$; thus, the interference deterioration is given by
    \begin{equation}
    I \sim 
    1 + I_0 \exp{\left(-\frac{\pi^2 N}{2500}\right)}.
    \end{equation}
    \\
    \textit{Intensity losses.--}
    The intensity losses on the mirrors and beam splitters used in our experiment are about $1\%$, which is acceptable.
    By far larger losses ($\approx 10\%$) are associated with the phase shifters.
    Nevertheless, the use of anti-reflective coating would reduce these losses to $1\%$.
    The signal intensity on the detector would be $0.99^N \approx \exp{\left( -N/100 \right)}$.

\begin{figure*}
\centering
    \includegraphics[width=0.24\linewidth]{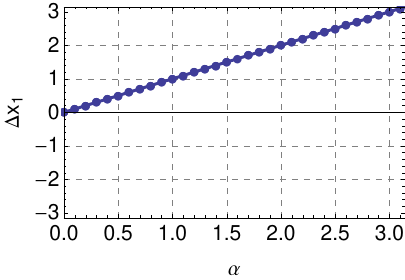}
    \includegraphics[width=0.24\linewidth]{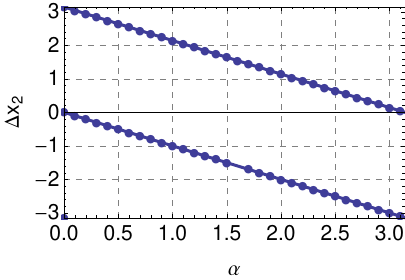}
    \includegraphics[width=0.24\linewidth]{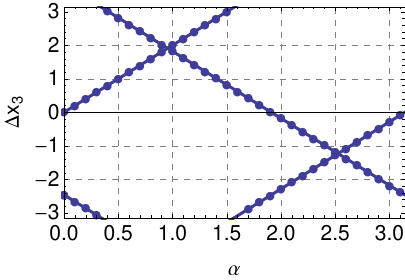}
    \includegraphics[width=0.24\linewidth]{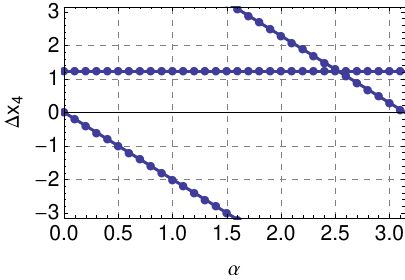}
    \caption{
        Numerically obtained parameters $\Delta\tilde{\textbf{x}}=[\Delta \tilde{x}_1, \Delta \tilde{x}_2, \Delta \tilde{x}_3, \Delta \tilde{x}_4]^T$ which for all $\phi$ satisfy the equations $ p_i(\Delta\tilde{\mathbf{x}}, \phi) = p_i(\Delta\mathbf{x}_0, \phi)$ as functions of $\alpha$.
        Here we parameterize $\Delta\mathbf{x}_0(\alpha)=[\alpha,-\alpha,2\alpha,-2\alpha]^T$.
        One can see that the same probability distributions $p_i(\phi)$ ($i=\{0,1,2\}$) can be realized with two different sets of parameters. 
        Note that by different sets we do not imply those which components differ by $2\pi$.}
    \label{sym_plot}
\end{figure*}

\subsection*{Fitting problem}
\label{ap_fit}
\subsubsection*{Overview of the problem}

    In this section, we shall discuss numerical methods which we employ for the analysis of experimental data.
    
    The measurable data consists of four real numbers: the geometrical angle of the swivel platform and three intensities.
    The fitting curve is determined by Eq.\,(14) of the main text and depends on the phase vector $\Delta\mathbf{x}=[x_1-x_1^F,x_2-x_2^F,x_3-x_3^F,x_4-x_4^F]^T$ along with the scaling and shifting parameters $\mathbf{a}=[a_1,a_2,a_3]^T$, $\mathbf{b}=[b_1,b_2,b_3]^T$, $\kappa$ and $\mu$.
    To fit the experimental data, we apply the method of least squares and search for the minimum of the following loss function:
    \begin{equation}
    \label{lossf}
         \mathcal{L}(\Delta\mathbf{x}, \mathbf{a}, \mathbf{b}, \kappa, \mu) =    \sum_{i=1}^{3} 
         \sum_{j=1}^N
        (p_i(\Delta\mathbf{x}, \kappa \cdot \phi_j + \mu, a_i, b_i) - \mathcal{P}_i(\phi_j))^2,
    \end{equation}
    where $\{\mathcal{P}_i(\phi_j)\}_{j=1}^N$ ($i=\{0,1,2\}$) is the array of the intensities experimentally measured by $i$th detector with the suffix $j$ representing the number of a data point.
    As the loss function appears to have multiple local minima, for the purpose of minimization we use the simulated annealing algorithm.
    
    The fitting error is determined by the maximum size of the neighbourhood $\mathcal{O}^F$ of $\mathbf{x}^F$ such that for any $\tilde{\mathbf{x}} \in \mathcal{O}^F$ the standard deviation of $p_i(\tilde{\mathbf{x}},\,\phi)$ from $p_i(\mathbf{x}^F,\,\phi)$ ($i\in\{0,\,1,\,2\}$) does not exceed the experimental error.
    
\subsubsection*{Loss landscape}
    The considered minimization problem has multiple solutions.
    For instance, Eqs.\,(11) of the main text and (\ref{lossf}) of SI show that the loss function $\mathcal{L}$ is periodical: 
    \begin{equation}
        \mathcal{L}(\Delta\mathbf{x}, \mathbf{a}, \mathbf{b}, \kappa, \mu) = \mathcal{L}(\Delta\mathbf{x}+2\pi\mathbf{n}, \mathbf{a}, \mathbf{b}, \kappa, \mu)
    \end{equation}
    for all integer-valued vectors $\mathbf{n}=[n_1,n_2,n_3,n_4]^T$.
    In order to find a proper solution of the fitting problem, one also needs to be aware of other patterns concerning the loss function.
    
    The study the landscape of the loss function lies in finding all possible symmetries $S: \mathbb{R}^4 \rightarrow \mathbb{R}^4$ such that for all $\Delta\mathbf{x}_0$ and $\phi$ one has 
    \begin{equation}
    \label{symeq}
        p_i(S(\Delta\mathbf{x}_0), \kappa \cdot \phi + \mu, a_i, b_i) = p_i(\Delta\mathbf{x}_0, \kappa \cdot \phi + \mu, a_i, b_i),
    \end{equation}
    with $i=\{0,1,2\}$.
    For our investigation, we conduct a series of numerical experiments in which we put $\kappa=1$, $\mu=0$, $a_i=1$, $b_i=0$ ($i=\{0,1,2\}$) (the role of the auxiliary scaling parameters is non-important).
    In each experiment we parameterize $\Delta\mathbf{x}_0$ using one parameter $\alpha$ and, with $\alpha$ assuming different values, we search for $\Delta\tilde{\mathbf{x}}$ satisfying the relation 
    \begin{equation}
    \label{pp}
        p_i(\Delta\tilde{\mathbf{x}}, \phi) \equiv p_i(\Delta\mathbf{x}_0, \phi).
    \end{equation}
    We then plot elements of $\Delta\tilde{\mathbf{x}}$ as functions of $\alpha$ and identify possible symmetries corresponding to Eq.\,(\ref{symeq}).
    For each experiment the parameterization is chosen differently, which ensures that none of the symmetries are missed out.
    
    Based on the obtained numerical data, we may suggest that aside from the trivial case of $2\pi$-periodicity there also exist only one linear symmetry given by 
\begin{equation}
\label{sym}
    S\begin{bmatrix}
           x_{1} \\
           x_{2} \\
           x_{3} \\
           x_{4} \\
         \end{bmatrix} = \begin{bmatrix}
           x_{1} \\
           x_{2} + \pi \\
           -x_{3} + 4 \arctan(\sqrt{2}) \\
           x_{3}  + x_4 + \pi-2 \arctan(\sqrt2) \\
         \end{bmatrix}.
\end{equation}
    Note that $S(S(\Delta\mathbf{x}))=\Delta\mathbf{x}+[0,2\pi,0,2\pi]^T$.
    The symmetry can be particularly seen in Fig.\,\ref{sym_plot} in which we put $\Delta\mathbf{x}_0(\alpha)=[\alpha,-\alpha,2\alpha,-2\alpha]^T$.
    The vertical axis corresponds to the elements of $\Delta\tilde{\mathbf{x}}$ satisfying Eq.\,(\ref{pp}), whereas the abscissa is the parameter $\alpha$.  
    The manifestation of the symmetry $S$ can be observed in the lines additional to those which correspond to $\Delta\mathbf{x}_0(\alpha)$.
    
    Assuming that the alignment process is relatively precise, the knowledge of symmetry gives us additional leverage for solving the fitting problem.
    After numerically finding the global minimum of $\mathcal{L}$ with the corresponding parameter vector $\Delta\mathbf{x}$, we can check whether there exists another one with the parameters $\Delta\mathbf{x}^\prime$ such that $|\Delta\mathbf{x}^\prime|<|\Delta\mathbf{x}|$. 
    This can be done by comparing $\Delta\mathbf{x}$ with $\Delta\mathbf{x}^\prime=S(\Delta\mathbf{x})+2\pi\mathbf{n}$ for various integer-valued vectors $\mathbf{n}$. 
    As a result, we obtain fitting parameters which are, considerably, the most consistent with the intended Fourier transform.


\begin{thebibliography}{10}
\urlstyle{rm}
\expandafter\ifx\csname url\endcsname\relax
  \def\url#1{\texttt{#1}}\fi
\expandafter\ifx\csname urlprefix\endcsname\relax\def\urlprefix{URL }\fi
\expandafter\ifx\csname doiprefix\endcsname\relax\def\doiprefix{DOI: }\fi
\providecommand{\bibinfo}[2]{#2}
\providecommand{\eprint}[2][]{\url{#2}}

\bibitem{Devoret}
\bibinfo{author}{Devoret, M.~H.} \& \bibinfo{author}{Schoelkopf, R.~J.}
\newblock \bibinfo{journal}{\bibinfo{title}{Superconducting circuits for
  quantum information: An outlook}}.
\newblock {\emph{\JournalTitle{Science}}} \textbf{\bibinfo{volume}{339}},
  \bibinfo{pages}{1169--1174}, \doiprefix\url{10.1126/science.1231930}
  (\bibinfo{year}{2013}).
\newblock
  \eprint{https://science.sciencemag.org/content/339/6124/1169.full.pdf}.

\bibitem{Cirac}
\bibinfo{author}{Cirac, J.~I.} \& \bibinfo{author}{Zoller, P.}
\newblock \bibinfo{journal}{\bibinfo{title}{Quantum computations with cold
  trapped ions}}.
\newblock {\emph{\JournalTitle{Phys. Rev. Lett.}}}
  \textbf{\bibinfo{volume}{74}}, \bibinfo{pages}{4091--4094},
  \doiprefix\url{10.1103/PhysRevLett.74.4091} (\bibinfo{year}{1995}).

\bibitem{Kane}
\bibinfo{author}{Kane, B.~E.}
\newblock \bibinfo{journal}{\bibinfo{title}{A silicon-based nuclear spin
  quantum computer}}.
\newblock {\emph{\JournalTitle{Nature}}} \textbf{\bibinfo{volume}{393}},
  \bibinfo{pages}{133--137}, \doiprefix\url{10.1038/30156}
  (\bibinfo{year}{1998}).

\bibitem{Loss}
\bibinfo{author}{Loss, D.} \& \bibinfo{author}{DiVincenzo, D.~P.}
\newblock \bibinfo{journal}{\bibinfo{title}{Quantum computation with quantum
  dots}}.
\newblock {\emph{\JournalTitle{Physical Review A}}}
  \textbf{\bibinfo{volume}{57}}, \bibinfo{pages}{120} (\bibinfo{year}{1998}).

\bibitem{Reck}
\bibinfo{author}{Reck, M.}, \bibinfo{author}{Zeilinger, A.},
  \bibinfo{author}{Bernstein, H.~J.} \& \bibinfo{author}{Bertani, P.}
\newblock \bibinfo{journal}{\bibinfo{title}{Experimental realization of any
  discrete unitary operator}}.
\newblock {\emph{\JournalTitle{Phys. Rev. Lett.}}}
  \textbf{\bibinfo{volume}{73}}, \bibinfo{pages}{58--61},
  \doiprefix\url{10.1103/PhysRevLett.73.58} (\bibinfo{year}{1994}).

\bibitem{Clauser}
\bibinfo{author}{Clauser, J.~F.} \& \bibinfo{author}{Dowling, J.~P.}
\newblock \bibinfo{journal}{\bibinfo{title}{Factoring integers with young's
  n-slit interferometer}}.
\newblock {\emph{\JournalTitle{Phys. Rev. A}}} \textbf{\bibinfo{volume}{53}},
  \bibinfo{pages}{4587--4590}, \doiprefix\url{10.1103/PhysRevA.53.4587}
  (\bibinfo{year}{1996}).

\bibitem{Knill}
\bibinfo{author}{Knill, E.}, \bibinfo{author}{Laflamme, R.} \&
  \bibinfo{author}{Milburn, G.~J.}
\newblock \bibinfo{journal}{\bibinfo{title}{A scheme for efficient quantum
  computation with linear optics}}.
\newblock {\emph{\JournalTitle{Nature}}} \textbf{\bibinfo{volume}{409}},
  \bibinfo{pages}{46} (\bibinfo{year}{2001}).

\bibitem{Tabia}
\bibinfo{author}{Tabia, G. N.~M.}
\newblock \bibinfo{journal}{\bibinfo{title}{Recursive multiport schemes for
  implementing quantum algorithms with photonic integrated circuits}}.
\newblock {\emph{\JournalTitle{Physical Review A}}}
  \textbf{\bibinfo{volume}{93}}, \bibinfo{pages}{012323}
  (\bibinfo{year}{2016}).

\bibitem{Poland}
\bibinfo{author}{Demkowicz-Dobrza{\'n}ski, R.}, \bibinfo{author}{Jarzyna, M.}
  \& \bibinfo{author}{Ko{\l}ody{\'n}ski, J.}
\newblock \bibinfo{title}{Quantum limits in optical interferometry}.
\newblock In \emph{\bibinfo{booktitle}{Progress in Optics}},
  vol.~\bibinfo{volume}{60}, \bibinfo{pages}{345--435}
  (\bibinfo{publisher}{Elsevier}, \bibinfo{year}{2015}).

\bibitem{O'Brien}
\bibinfo{author}{O'brien, J.~L.}
\newblock \bibinfo{journal}{\bibinfo{title}{Optical quantum computing}}.
\newblock {\emph{\JournalTitle{Science}}} \textbf{\bibinfo{volume}{318}},
  \bibinfo{pages}{1567--1570} (\bibinfo{year}{2007}).

\bibitem{Dowling}
\bibinfo{author}{Dowling, J.~P.} \& \bibinfo{author}{Seshadreesan, K.~P.}
\newblock \bibinfo{journal}{\bibinfo{title}{Quantum optical technologies for
  metrology, sensing, and imaging}}.
\newblock {\emph{\JournalTitle{Journal of Lightwave Technology}}}
  \textbf{\bibinfo{volume}{33}}, \bibinfo{pages}{2359--2370}
  (\bibinfo{year}{2015}).

\bibitem{Carolan}
\bibinfo{author}{Carolan, J.} \emph{et~al.}
\newblock \bibinfo{journal}{\bibinfo{title}{Universal linear optics}}.
\newblock {\emph{\JournalTitle{Science}}} \textbf{\bibinfo{volume}{349}},
  \bibinfo{pages}{711--716}, \doiprefix\url{10.1126/science.aab3642}
  (\bibinfo{year}{2015}).
\newblock
  \eprint{https://science.sciencemag.org/content/349/6249/711.full.pdf}.

\bibitem{Ralph}
\bibinfo{author}{Ralph, T.~C.}, \bibinfo{author}{Gilchrist, A.},
  \bibinfo{author}{Milburn, G.~J.}, \bibinfo{author}{Munro, W.~J.} \&
  \bibinfo{author}{Glancy, S.}
\newblock \bibinfo{journal}{\bibinfo{title}{Quantum computation with optical
  coherent states}}.
\newblock {\emph{\JournalTitle{Physical Review A}}}
  \textbf{\bibinfo{volume}{68}}, \bibinfo{pages}{042319}
  (\bibinfo{year}{2003}).

\bibitem{Kwiat}
\bibinfo{author}{Cerf, N.~J.}, \bibinfo{author}{Adami, C.} \&
  \bibinfo{author}{Kwiat, P.~G.}
\newblock \bibinfo{journal}{\bibinfo{title}{Optical simulation of quantum
  logic}}.
\newblock {\emph{\JournalTitle{Physical Review A}}}
  \textbf{\bibinfo{volume}{57}}, \bibinfo{pages}{R1477} (\bibinfo{year}{1998}).

\bibitem{Lloyd}
\bibinfo{author}{Lloyd, S.} \& \bibinfo{author}{Braunstein, S.~L.}
\newblock \bibinfo{journal}{\bibinfo{title}{Quantum computation over continuous
  variables}}.
\newblock {\emph{\JournalTitle{Phys. Rev. Lett.}}}
  \textbf{\bibinfo{volume}{82}}, \bibinfo{pages}{1784--1787},
  \doiprefix\url{10.1103/PhysRevLett.82.1784} (\bibinfo{year}{1999}).

\bibitem{Daryanoosh}
\bibinfo{author}{Daryanoosh, S.}, \bibinfo{author}{Slussarenko, S.},
  \bibinfo{author}{Berry, D.~W.}, \bibinfo{author}{Wiseman, H.~M.} \&
  \bibinfo{author}{Pryde, G.~J.}
\newblock \bibinfo{journal}{\bibinfo{title}{Experimental optical phase
  measurement approaching the exact heisenberg limit}}.
\newblock {\emph{\JournalTitle{Nature Communications}}}
  \textbf{\bibinfo{volume}{9}}, \bibinfo{pages}{4606} (\bibinfo{year}{2018}).

\bibitem{Su2017}
\bibinfo{author}{Su, Z.-E.} \emph{et~al.}
\newblock \bibinfo{journal}{\bibinfo{title}{Multiphoton interference in quantum
  fourier transform circuits and applications to quantum metrology}}.
\newblock {\emph{\JournalTitle{Phys. Rev. Lett.}}}
  \textbf{\bibinfo{volume}{119}}, \bibinfo{pages}{080502},
  \doiprefix\url{10.1103/PhysRevLett.119.080502} (\bibinfo{year}{2017}).

\bibitem{Clements}
\bibinfo{author}{Clements, W.~R.}, \bibinfo{author}{Humphreys, P.~C.},
  \bibinfo{author}{Metcalf, B.~J.}, \bibinfo{author}{Kolthammer, W.~S.} \&
  \bibinfo{author}{Walmsley, I.~A.}
\newblock \bibinfo{journal}{\bibinfo{title}{Optimal design for universal
  multiport interferometers}}.
\newblock {\emph{\JournalTitle{Optica}}} \textbf{\bibinfo{volume}{3}},
  \bibinfo{pages}{1460--1465}, \doiprefix\url{10.1364/OPTICA.3.001460}
  (\bibinfo{year}{2016}).

\bibitem{Su}
\bibinfo{author}{Su, D.} \emph{et~al.}
\newblock \bibinfo{journal}{\bibinfo{title}{Implementing quantum algorithms on
  temporal photonic cluster states}}.
\newblock {\emph{\JournalTitle{Phys. Rev. A}}} \textbf{\bibinfo{volume}{98}},
  \bibinfo{pages}{032316}, \doiprefix\url{10.1103/PhysRevA.98.032316}
  (\bibinfo{year}{2018}).

\bibitem{Kiloran}
\bibinfo{author}{Killoran, N.} \emph{et~al.}
\newblock \bibinfo{journal}{\bibinfo{title}{Continuous-variable quantum neural
  networks}}.
\newblock {\emph{\JournalTitle{Phys. Rev. Research}}}
  \textbf{\bibinfo{volume}{1}}, \bibinfo{pages}{033063},
  \doiprefix\url{10.1103/PhysRevResearch.1.033063} (\bibinfo{year}{2019}).

\bibitem{Suslov}
\bibinfo{author}{Suslov, M.~V.}, \bibinfo{author}{Lesovik, G.~B.} \&
  \bibinfo{author}{Blatter, G.}
\newblock \bibinfo{journal}{\bibinfo{title}{Quantum abacus for counting and
  factorizing numbers}}.
\newblock {\emph{\JournalTitle{Phys. Rev. A}}} \textbf{\bibinfo{volume}{83}},
  \bibinfo{pages}{052317}, \doiprefix\url{10.1103/PhysRevA.83.052317}
  (\bibinfo{year}{2011}).

\bibitem{Danilin}
\bibinfo{author}{Danilin, S.} \emph{et~al.}
\newblock \bibinfo{journal}{\bibinfo{title}{Quantum-enhanced magnetometry by
  phase estimation algorithms with a single artificial atom}}.
\newblock {\emph{\JournalTitle{npj Quantum Information}}}
  \textbf{\bibinfo{volume}{4}}, \bibinfo{pages}{29},
  \doiprefix\url{10.1038/s41534-018-0078-y} (\bibinfo{year}{2018}).

\bibitem{Shlyakhov}
\bibinfo{author}{Shlyakhov, A.~R.} \emph{et~al.}
\newblock \bibinfo{journal}{\bibinfo{title}{Quantum metrology with a transmon
  qutrit}}.
\newblock {\emph{\JournalTitle{Phys. Rev. A}}} \textbf{\bibinfo{volume}{97}},
  \bibinfo{pages}{022115}, \doiprefix\url{10.1103/PhysRevA.97.022115}
  (\bibinfo{year}{2018}).

\bibitem{Qi2018}
\bibinfo{author}{Qi, H.}, \bibinfo{author}{Helt, L.~G.}, \bibinfo{author}{Su,
  D.}, \bibinfo{author}{Vernon, Z.} \& \bibinfo{author}{Brádler, K.}
\newblock \bibinfo{title}{Linear multiport photonic interferometers: loss
  analysis of temporally-encoded architectures} (\bibinfo{year}{2018}).
\newblock \eprint{1812.07015}.

\bibitem{Guise}
\bibinfo{author}{de~Guise, H.}, \bibinfo{author}{Di~Matteo, O.} \&
  \bibinfo{author}{S\'anchez-Soto, L.~L.}
\newblock \bibinfo{journal}{\bibinfo{title}{Simple factorization of unitary
  transformations}}.
\newblock {\emph{\JournalTitle{Phys. Rev. A}}} \textbf{\bibinfo{volume}{97}},
  \bibinfo{pages}{022328}, \doiprefix\url{10.1103/PhysRevA.97.022328}
  (\bibinfo{year}{2018}).

\bibitem{Harris}
\bibinfo{author}{Harris, N.~C.} \emph{et~al.}
\newblock \bibinfo{journal}{\bibinfo{title}{Large-scale quantum photonic
  circuits in silicon}}.
\newblock {\emph{\JournalTitle{Nanophotonics}}} \textbf{\bibinfo{volume}{5}},
  \bibinfo{pages}{456--468}, \doiprefix\url{10.1515/nanoph-2015-0146}
  (\bibinfo{year}{2016}).

\end{thebibliography}


\begin{thebibliography}{10}
\urlstyle{rm}
\expandafter\ifx\csname url\endcsname\relax
  \def\url#1{\texttt{#1}}\fi
\expandafter\ifx\csname urlprefix\endcsname\relax\def\urlprefix{URL }\fi
\expandafter\ifx\csname doiprefix\endcsname\relax\def\doiprefix{DOI: }\fi
\providecommand{\bibinfo}[2]{#2}
\providecommand{\eprint}[2][]{\url{#2}}

\bibitem{Shor1994}
\bibinfo{author}{Shor, P.}
\newblock \bibinfo{title}{Algorithms for quantum computation: discrete
  logarithms and factoring}.
\newblock \emph{\bibinfo{booktitle}{Proceedings 35th Annual Symposium on
  Foundations of Computer Science}} 
  (\bibinfo{publisher}{{IEEE} Comput. Soc. Press}).

\bibitem{Lloyd2007}
\bibinfo{author}{Harrow, A.~W.}, \bibinfo{author}{Hassidim, A.} \&
  \bibinfo{author}{Lloyd, S.}
\newblock \bibinfo{journal}{\bibinfo{title}{Quantum algorithm for linear
  systems of equations}}.
\newblock {\emph{\JournalTitle{Physical Review Letters}}}
  \textbf{\bibinfo{volume}{103}}
  (\bibinfo{year}{2009}).

\bibitem{Peruzzo2014}
\bibinfo{author}{Peruzzo, A.} \emph{et~al.}
\newblock \bibinfo{journal}{\bibinfo{title}{A variational eigenvalue solver on
  a photonic quantum processor}}.
\newblock {\emph{\JournalTitle{Nature Communications}}}
  \textbf{\bibinfo{volume}{5}} 
  (\bibinfo{year}{2014}).

\bibitem{NielsenChuang}
\bibinfo{author}{Nielsen, M.~A.} \& \bibinfo{author}{Chuang, I.}
\newblock \emph{\bibinfo{title}{Quantum Computation and Quantum Information:
  10th Anniversary Edition}} (\bibinfo{publisher}{Cambridge University Press},
  \bibinfo{year}{2011}).

\bibitem{Giovannetti2004}
\bibinfo{author}{Giovannetti, V.}
\newblock \bibinfo{journal}{\bibinfo{title}{Quantum-enhanced measurements:
  Beating the standard quantum limit}}.
\newblock {\emph{\JournalTitle{Science}}} \textbf{\bibinfo{volume}{306}},
  \bibinfo{pages}{1330--1336} 
  (\bibinfo{year}{2004}).

\bibitem{Lesovik2010}
\bibinfo{author}{Lesovik, G.~B.}, \bibinfo{author}{Suslov, M.~V.} \&
  \bibinfo{author}{Blatter, G.}
\newblock \bibinfo{journal}{\bibinfo{title}{Quantum counting algorithm and its
  application in mesoscopic physics}}.
\newblock {\emph{\JournalTitle{Physical Review A}}}
  \textbf{\bibinfo{volume}{82}} 
  (\bibinfo{year}{2010}).

\bibitem{Suslov2011}
\bibinfo{author}{Suslov, M.~V.}, \bibinfo{author}{Lesovik, G.~B.} \&
  \bibinfo{author}{Blatter, G.}
\newblock \bibinfo{journal}{\bibinfo{title}{Quantum abacus for counting and
  factorizing numbers}}.
\newblock {\emph{\JournalTitle{Phys. Rev. A}}} \textbf{\bibinfo{volume}{83}},
  \bibinfo{pages}{052317} 
  (\bibinfo{year}{2011}).

\bibitem{Giovannetti2011}
\bibinfo{author}{Giovannetti, V.}, \bibinfo{author}{Lloyd, S.} \&
  \bibinfo{author}{Maccone, L.}
\newblock \bibinfo{journal}{\bibinfo{title}{Advances in quantum metrology}}.
\newblock {\emph{\JournalTitle{Nature Photonics}}}
  \textbf{\bibinfo{volume}{5}}, \bibinfo{pages}{222--229}
  (\bibinfo{year}{2011}).

\bibitem{Degen2017}
\bibinfo{author}{Degen, C.}, \bibinfo{author}{Reinhard, F.} \&
  \bibinfo{author}{Cappellaro, P.}
\newblock \bibinfo{journal}{\bibinfo{title}{Quantum sensing}}.
\newblock {\emph{\JournalTitle{Reviews of Modern Physics}}}
  \textbf{\bibinfo{volume}{89}} 
  (\bibinfo{year}{2017}).

\bibitem{Pirandola2018}
\bibinfo{author}{Pirandola, S.}, \bibinfo{author}{Bardhan, B.~R.},
  \bibinfo{author}{Gehring, T.}, \bibinfo{author}{Weedbrook, C.} \&
  \bibinfo{author}{Lloyd, S.}
\newblock \bibinfo{journal}{\bibinfo{title}{Advances in photonic quantum
  sensing}}.
\newblock {\emph{\JournalTitle{Nature Photonics}}}
  \textbf{\bibinfo{volume}{12}}, \bibinfo{pages}{724--733}
  (\bibinfo{year}{2018}).

\bibitem{Waldherr2011}
\bibinfo{author}{Waldherr, G.} \emph{et~al.}
\newblock \bibinfo{journal}{\bibinfo{title}{High-dynamic-range magnetometry
  with a single nuclear spin in diamond}}.
\newblock {\emph{\JournalTitle{Nature Nanotechnology}}}
  \textbf{\bibinfo{volume}{7}}, \bibinfo{pages}{105--108}
  (\bibinfo{year}{2011}).

\bibitem{Bal2012}
\bibinfo{author}{Bal, M.}, \bibinfo{author}{Deng, C.},
  \bibinfo{author}{Orgiazzi, J.-L.}, \bibinfo{author}{Ong, F.} \&
  \bibinfo{author}{Lupascu, A.}
\newblock \bibinfo{journal}{\bibinfo{title}{Ultrasensitive magnetic field
  detection using a single artificial atom}}.
\newblock {\emph{\JournalTitle{Nature Communications}}}
  \textbf{\bibinfo{volume}{3}} 
  (\bibinfo{year}{2012}).

\bibitem{Puentes2014}
\bibinfo{author}{Puentes, G.}, \bibinfo{author}{Waldherr, G.},
  \bibinfo{author}{Neumann, P.}, \bibinfo{author}{Balasubramanian, G.} \&
  \bibinfo{author}{Wrachtrup, J.}
\newblock \bibinfo{journal}{\bibinfo{title}{Efficient route to high-bandwidth
  nanoscale magnetometry using single spins in diamond}}.
\newblock {\emph{\JournalTitle{Sci. Rep.}}}
  \textbf{\bibinfo{volume}{4}} 
  (\bibinfo{year}{2014}).

\bibitem{Bonato2015}
\bibinfo{author}{Bonato, C.} \emph{et~al.}
\newblock \bibinfo{journal}{\bibinfo{title}{Optimized quantum sensing with a
  single electron spin using real-time adaptive measurements}}.
\newblock {\emph{\JournalTitle{Nature Nanotechnology}}}
  \textbf{\bibinfo{volume}{11}}, \bibinfo{pages}{247--252}
  (\bibinfo{year}{2015}).

\bibitem{Chen2017}
\bibinfo{author}{Chen, M.} \emph{et~al.}
\newblock \bibinfo{journal}{\bibinfo{title}{Quantum metrology with single spins
  in diamond under ambient conditions}}.
\newblock {\emph{\JournalTitle{National Science Review}}}
  \textbf{\bibinfo{volume}{5}}, \bibinfo{pages}{346--355}
  (\bibinfo{year}{2017}).

\bibitem{Danilin}
\bibinfo{author}{Danilin, S.} \emph{et~al.}
\newblock \bibinfo{journal}{\bibinfo{title}{Quantum-enhanced magnetometry by
  phase estimation algorithms with a single artificial atom}}.
\newblock {\emph{\JournalTitle{npj Quantum Information}}}
  \textbf{\bibinfo{volume}{4}}, \bibinfo{pages}{29}
  (\bibinfo{year}{2018}).

\bibitem{Shlyakhov}
\bibinfo{author}{Shlyakhov, A.~R.} \emph{et~al.}
\newblock \bibinfo{journal}{\bibinfo{title}{Quantum metrology with a transmon
  qutrit}}.
\newblock {\emph{\JournalTitle{Phys. Rev. A}}} \textbf{\bibinfo{volume}{97}},
  \bibinfo{pages}{022115} 
  (\bibinfo{year}{2018}).

\bibitem{Higgins2007}
\bibinfo{author}{Higgins, B.~L.}, \bibinfo{author}{Berry, D.~W.},
  \bibinfo{author}{Bartlett, S.~D.}, \bibinfo{author}{Wiseman, H.~M.} \&
  \bibinfo{author}{Pryde, G.~J.}
\newblock \bibinfo{journal}{\bibinfo{title}{Entanglement-free
  heisenberg-limited phase estimation}}.
\newblock {\emph{\JournalTitle{Nature}}} \textbf{\bibinfo{volume}{450}},
  \bibinfo{pages}{393--396} 
  (\bibinfo{year}{2007}).

\bibitem{Knill}
\bibinfo{author}{Knill, E.}, \bibinfo{author}{Laflamme, R.} \&
  \bibinfo{author}{Milburn, G.~J.}
\newblock \bibinfo{journal}{\bibinfo{title}{A scheme for efficient quantum
  computation with linear optics}}.
\newblock {\emph{\JournalTitle{Nature}}} \textbf{\bibinfo{volume}{409}},
  \bibinfo{pages}{46} (\bibinfo{year}{2001}).

\bibitem{Poland}
\bibinfo{author}{Demkowicz-Dobrza{\'n}ski, R.}, \bibinfo{author}{Jarzyna, M.}
  \& \bibinfo{author}{Ko{\l}ody{\'n}ski, J.}
\newblock \bibinfo{title}{Quantum limits in optical interferometry}.
\newblock \emph{\bibinfo{booktitle}{Prog. Optics}},
  vol.~\bibinfo{volume}{60}, \bibinfo{pages}{345--435}
  (\bibinfo{year}{2015}).

\bibitem{Dowling}
\bibinfo{author}{Dowling, J.~P.} \& \bibinfo{author}{Seshadreesan, K.~P.}
\newblock \bibinfo{journal}{\bibinfo{title}{Quantum optical technologies for
  metrology, sensing, and imaging}}.
\newblock {\emph{\JournalTitle{Journal of Lightwave Technology}}}
  \textbf{\bibinfo{volume}{33}}, \bibinfo{pages}{2359--2370}
  (\bibinfo{year}{2015}).

\bibitem{Carolan}
\bibinfo{author}{Carolan, J.} \emph{et~al.}
\newblock \bibinfo{journal}{\bibinfo{title}{Universal linear optics}}.
\newblock {\emph{\JournalTitle{Science}}} \textbf{\bibinfo{volume}{349}},
  \bibinfo{pages}{711--716} 
  (\bibinfo{year}{2015}).
\newblock

\bibitem{Tan2019}
\bibinfo{author}{Tan, S.-H.} \& \bibinfo{author}{Rohde, P.~P.}
\newblock \bibinfo{journal}{\bibinfo{title}{The resurgence of the linear optics
  quantum interferometer {\textemdash} recent advances {\&} applications}}.
\newblock {\emph{\JournalTitle{Reviews in Physics}}}
  \textbf{\bibinfo{volume}{4}}, \bibinfo{pages}{100030}
  (\bibinfo{year}{2019}).
  
\bibitem{LIGO}
Abbott, B. P. \textit{et al.} LIGO: the Laser Interferometer Gravitational-Wave Observatory. Rep. Prog. Phys. \textbf{72}, 076901 (2009).

\bibitem{Reck}
\bibinfo{author}{Reck, M.}, \bibinfo{author}{Zeilinger, A.},
  \bibinfo{author}{Bernstein, H.~J.} \& \bibinfo{author}{Bertani, P.}
\newblock \bibinfo{journal}{\bibinfo{title}{Experimental realization of any
  discrete unitary operator}}.
\newblock {\emph{\JournalTitle{Phys. Rev. Lett.}}}
  \textbf{\bibinfo{volume}{73}}, \bibinfo{pages}{58--61}
  (\bibinfo{year}{1994}).

\bibitem{Clements}
\bibinfo{author}{Clements, W.~R.}, \bibinfo{author}{Humphreys, P.~C.},
  \bibinfo{author}{Metcalf, B.~J.}, \bibinfo{author}{Kolthammer, W.~S.} \&
  \bibinfo{author}{Walmsley, I.~A.}
\newblock \bibinfo{journal}{\bibinfo{title}{Optimal design for universal
  multiport interferometers}}.
\newblock {\emph{\JournalTitle{Optica}}} \textbf{\bibinfo{volume}{3}},
  \bibinfo{pages}{1460--1465} 
  (\bibinfo{year}{2016}).

\bibitem{Qi2018}
\bibinfo{author}{Qi, H.}, \bibinfo{author}{Helt, L.~G.}, \bibinfo{author}{Su,
  D.}, \bibinfo{author}{Vernon, Z.} \& \bibinfo{author}{Brádler, K.}
\newblock \bibinfo{title}{Linear multiport photonic interferometers: loss
  analysis of temporally-encoded architectures}.
  Preprint at https://arxiv.org/abs/1812.07015 (\bibinfo{year}{2018}).

\bibitem{Guise}
\bibinfo{author}{de~Guise, H.}, \bibinfo{author}{Di~Matteo, O.} \&
  \bibinfo{author}{S\'anchez-Soto, L.~L.}
\newblock \bibinfo{journal}{\bibinfo{title}{Simple factorization of unitary
  transformations}}.
\newblock {\emph{\JournalTitle{Phys. Rev. A}}} \textbf{\bibinfo{volume}{97}},
  \bibinfo{pages}{022328} 
  (\bibinfo{year}{2018}).

\bibitem{Harris}
\bibinfo{author}{Harris, N.~C.} \emph{et~al.}
\newblock \bibinfo{journal}{\bibinfo{title}{Large-scale quantum photonic
  circuits in silicon}}.
\newblock {\emph{\JournalTitle{Nanophotonics}}} \textbf{\bibinfo{volume}{5}},
  \bibinfo{pages}{456--468} 
  (\bibinfo{year}{2016}).

\bibitem{Su2017}
\bibinfo{author}{Su, Z.-E.} \emph{et~al.}
\newblock \bibinfo{journal}{\bibinfo{title}{Multiphoton interference in quantum
  fourier transform circuits and applications to quantum metrology}}.
\newblock {\emph{\JournalTitle{Phys. Rev. Lett.}}}
  \textbf{\bibinfo{volume}{119}}, \bibinfo{pages}{080502}
  (\bibinfo{year}{2017}).

\bibitem{Daryanoosh}
\bibinfo{author}{Daryanoosh, S.}, \bibinfo{author}{Slussarenko, S.},
  \bibinfo{author}{Berry, D.~W.}, \bibinfo{author}{Wiseman, H.~M.} \&
  \bibinfo{author}{Pryde, G.~J.}
\newblock \bibinfo{journal}{\bibinfo{title}{Experimental optical phase
  measurement approaching the exact heisenberg limit}}.
\newblock {\emph{\JournalTitle{Nature Communications}}}
  \textbf{\bibinfo{volume}{9}}, \bibinfo{pages}{4606} (\bibinfo{year}{2018}).

\end{thebibliography}
\end{document}